\documentclass[lettersize,journal]{IEEEtran}

\IEEEoverridecommandlockouts                              

\usepackage{graphics} 
\usepackage{epsfig} 
\usepackage{amsmath} 

\usepackage{stfloats}
\usepackage{url}
\usepackage{verbatim}
\usepackage{graphicx}
\usepackage{cite}
\usepackage{amssymb}
\usepackage{booktabs}
\usepackage{makecell}
\usepackage{threeparttable}
\usepackage{subfigure}
\usepackage{multirow}
\usepackage{lscape}
\usepackage[T1]{fontenc}
\usepackage[figuresright]{rotating}

\title{\LARGE \bf
4D Millimeter-Wave Radar in Autonomous Driving: A Survey
}

\author{Zeyu Han$^{1}$, Jiahao Wang$^{1}$, Zikun Xu$^{1}$, Shuocheng Yang$^{2}$, Zhouwei Kong$^{3}$, Lei He$^{1}$, Shaobing Xu$^{1}$, \\Jianqiang Wang$^{1,*}$,  Keqiang Li$^{1,*}$
\thanks{Zeyu Han and Jiahao Wang contribute equally to this work.}
\thanks{This work was supported by the National Natural Science Foundation of China (NSFC) under grant number 52221005 and Tsinghua University - Chongqing Changan Automobile Co., Ltd. Joint Research Project.}
\thanks{$^{1}$School of Vehicle and Mobility, Tsinghua University, Beijing, China}
\thanks{$^{2}$Xingjian College, Tsinghua University, Beijing, China}
\thanks{$^{3}$Changan Automobile Co., Ltd., Chongqing, China}
\thanks{$^{*}$Correspondence: {\tt wjqlws@tsinghua.edu.cn} (J.W.), {\tt likq@tsinghua.edu.cn} (K.L.)}
}

\begin{document}

\maketitle

\begin{abstract}

The 4D millimeter-wave (mmWave) radar, proficient in measuring the range, azimuth, elevation, and velocity of targets, has attracted considerable interest within the autonomous driving community. This is attributed to its robustness in extreme environments and the velocity and elevation measurement capabilities. However, despite the rapid advancement in research related to its sensing theory and application, there is a conspicuous absence of comprehensive surveys on the subject of 4D mmWave radars. In an effort to bridge this gap and stimulate future research, this paper presents an exhaustive survey on the utilization of 4D mmWave radars in autonomous driving. Initially, the paper provides reviews on the theoretical background and progress of 4D mmWave radars, encompassing aspects such as the signal processing workflow, resolution improvement approaches, and extrinsic calibration process. Learning-based radar data quality improvement methods are present following. Then, this paper introduces relevant datasets and application algorithms in autonomous driving perception and localization tasks. Finally, this paper concludes by forecasting future trends in the realm of the 4D mmWave radar in autonomous driving. To the best of our knowledge, this is the first survey specifically dedicated to the 4D mmWave radar in autonomous driving.



\end{abstract}



\section{Introduction} \label{Sec:Intro}

Autonomous driving technology, which aspires to provide safe, convenient, and comfortable transportation experiences, is advancing at a remarkable pace. To realize high-level autonomous driving, the capabilities of environment perception and localization are indispensable. Consequently, the sensors deployed on autonomous vehicles, such as cameras, LiDARs, and radars, along with their application algorithms, are garnering increasing research interest. 

Among the various sensors, mmWave radars, with their acknowledged advantages of compact size, cost-effectiveness, all-weather adaptation, velocity-measuring capability, and long detection range, etc.\cite{jiang4DHighResolutionImagery2023}, have always been extensively employed in autonomous driving. However, conventional mmWave radars, often referred to as 3D mmWave radars, demonstrate limited efficacy in measuring the elevation of targets, and their data typically encompassing only range, azimuth, and Doppler velocity information. Additionally, 3D mmWave radars suffer from clutter, noise, and low resolution, particularly in the angular dimension.  These limitations further constrain their suitability for intricate perception tasks.

\begin{figure}
    \centering
    \includegraphics[width = \linewidth]{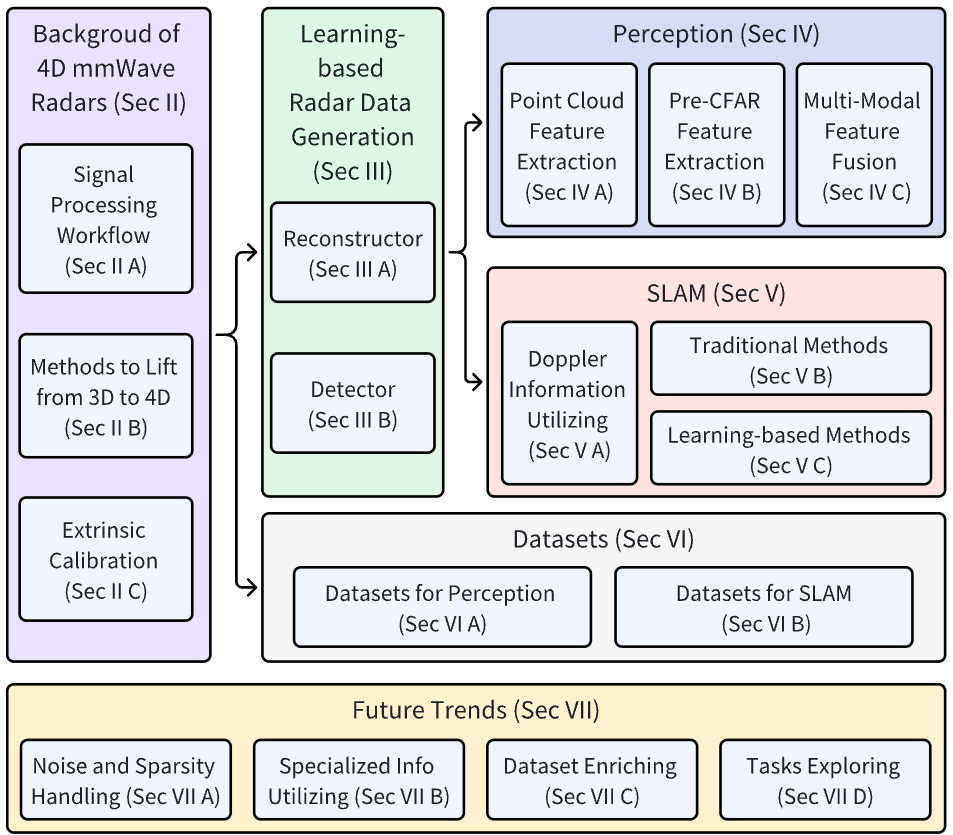}
    \caption{The main pipeline of this survey.}
    \label{Fig:Introduction}
\end{figure}

\begin{figure}[htbp]
    \begin{minipage}[b]{0.3\linewidth}
        \centering
        \includegraphics[width=\textwidth]{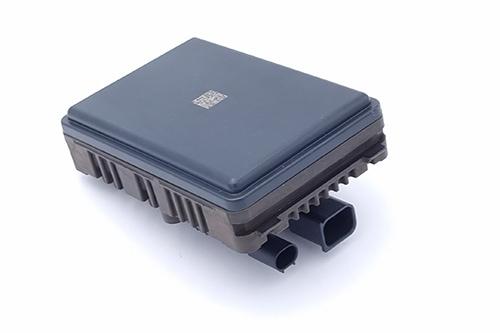}
        \centerline{(a)}
        
        \vspace{10pt} 
    
        \includegraphics[width=\textwidth]{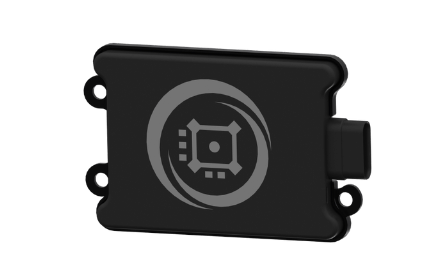}
        \centerline{(b)}
    \end{minipage}
    \hfill 
    \begin{minipage}[b]{0.7\linewidth}
        \centering
        \includegraphics[width=\textwidth]{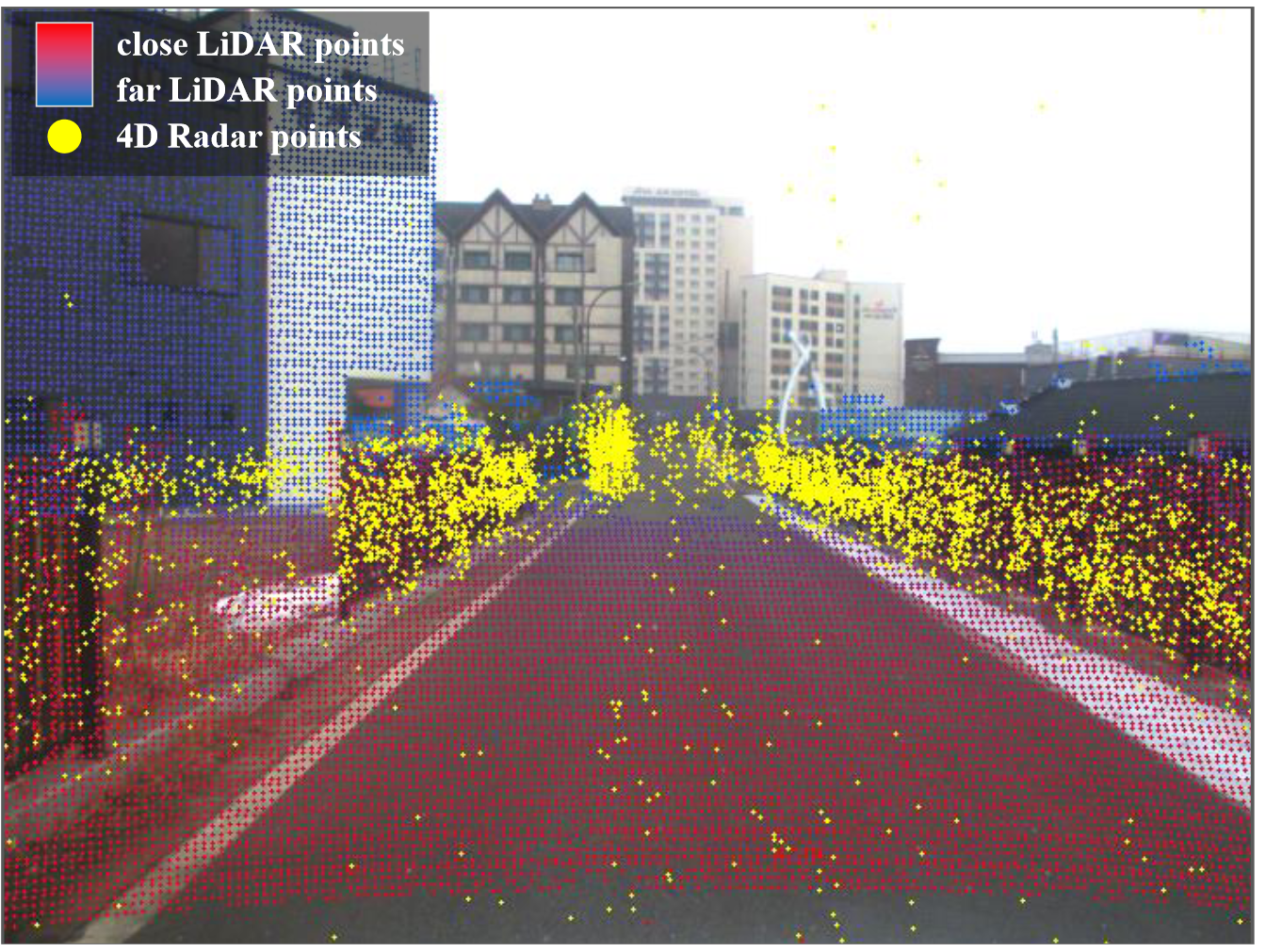}
        \centerline{(c)}
    \end{minipage}
\caption{The Continental ARS548RDI 4D mmWave radar (a), Oculii Eagle 4D mmWave radar (b) and the point cloud of Oculii Eagle comparing with the Ouster 128-channel LiDAR (c). \cite{choiMSCRAD4RROSBasedAutomotive2023}}
\label{Fig:oculii}
\end{figure}

The recent advancement of multiple-input multiple-output (MIMO) antenna technology has catalyzed a significant enhancement in elevational resolution, leading to the emergence of 4D mmWave radars. As the name suggests, 4D mmWave radars are capable of measuring four distinct types of target information: range, azimuth, elevation, and velocity. In addition to the augmented elevational resolution, 4D mmWave radars still preserve the salient advantages of their 3D predecessors. The comparison among autonomous driving sensors, including 4D mmWave radar, 3D mmWave radar, LiDAR, RGB camera and thermal camera is shown in Table. \ref{Tab:sensors}. The 4D mmWave radar distinctly holds advantages in velocity measurement, detection range, all-environment robustness and low cost. Enterprises that involve in the 4D mmWave radar industry ranging from conventional suppliers like Bosch, Continental, and ZF, to a host of burgeoning tech companies such as Arbe, Huawei, and Oculii. An illustrative example of this technology is the Oculii Eagle 4D mmWave radar, which, when compared with the Ouster 128-channel LiDAR, demonstrates its capabilities such as long detection range through the point cloud representation as depicted in Fig. \ref{Fig:oculii}.

\begin{table*}[!htbp]
    \centering
    \caption{Comparison among autonomous driving sensors and data formats}
    \resizebox{\linewidth}{!}{
    \begin{tabular}{cccccc} 
        \toprule
        Features &4D mmWave Radar &3D mmWave Radar &LiDAR &RGB Camera &Thermal Camera \\
        \midrule
        Range Resolution     &High &High &Very High &Low &Low \\
        Azimuth Resolution   &High &Moderate &Very High &Moderate &Moderate\\
        Elevation Resolution &High &Unmeasurable &Very High &Moderate &Moderate \\
        Velocity Resolution  &High &High &Unmeasurable &Unmeasurable &Unmeasurable \\
        Detection Range      &High &High &Moderate &Low &Moderate \\
        Surface Measurement  &Texture &Texture &No &Color &Thermal Signature \\
        Lighting Robustness  &High &High &High &Low &High \\
        Weather Robustness   &High &High &Low &Low &High \\
        Cost                 &Moderate &Low &High &Moderate &High \\
        \bottomrule
    \end{tabular}}
    \label{Tab:sensors}
\end{table*}

However, as a newly developed sensor, the 4D mmWave radar also presents some challenges due to its inherent characteristics. On the one hand, the raw data volume generated by the 4D mmWave radar substantially exceeds that of its traditional counterpart, thereby presenting formidable problems in signal processing and data generation. on the other hand, the sparsity and noise natural in 4D mmWave radar point clouds, generated in the existing signal processing workflow are notably more severe than those in LiDAR point clouds. Aiming at solving these issues, as well as utilizing 4D mmWave radar features such as Doppler and elevation measurement, a great number of researchers have engaged in studies within the fields of 4D mmWave radar-based data generation\cite{brodeskiDeepRadarDetector2019, chengNovelRadarPoint2022}, perception\cite{yanMVFANMultiViewFeature2023, liuSMURFSpatialMultiRepresentation2023} and SLAM(Simultaneous Localization and Mapping)\cite{zhuang4DIRIOM4D2023, zhuo4DRVONetDeep4D2023}.

In recent years, numerous surveys have been conducted on the theory and application of mmWave radars\cite{bilikRiseRadarAutonomous2019, venonMillimeterWaveFMCW2022, harlowNewWaveRobotics2023, zhouMMWRadarBasedTechnologies2020, weiMmWaveRadarVision2022, fan4DMmWaveRadar2024}, but most of them are centered on 3D mmWave radars. Bilik et al. \cite{bilikRiseRadarAutonomous2019} have reviewed the challenges faced by mmWave radars in autonomous driving and its future trends. Venon et al. \cite{venonMillimeterWaveFMCW2022} have provided a comprehensive summary of the theory and existing perception algorithms of mmWave radar in autonomous driving, while Harlow et al. \cite{harlowNewWaveRobotics2023} have concentrated on mmWave radar applications in robotics for their survey.
%


\begin{figure*}
    \centering
    \includegraphics[width = \linewidth]{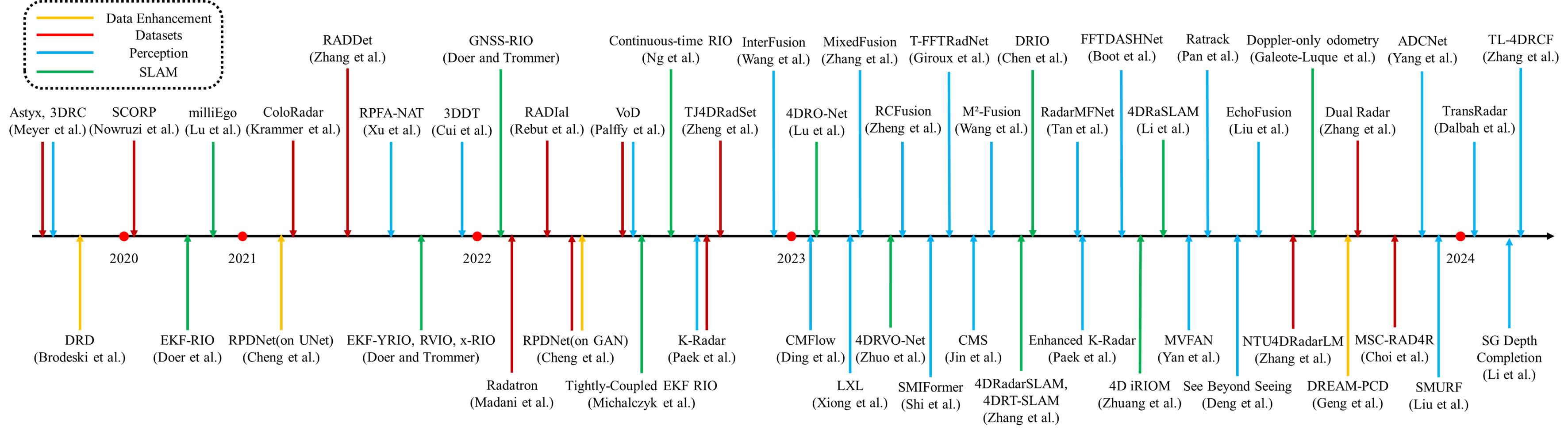}
    \caption{The timeline of 4D mmWave radar-related works, including learning-based radar data generation methods, perception and SLAM algorithms, and datasets}
    \label{Fig:timeline}
\end{figure*}

Despite the transformative emergence of 4D mmWave radars and associated algorithms, there have been few specialized surveys. Liu et al. \cite{liuWhichFrameworkSuitable2024} compare different pipelines of 4D mmWave radar-based object tracking algorithms. Fan et al. \cite{fan4DMmWaveRadar2024} summarize perception and SLAM applications for 4D mmWave radar, but ignore the quite important radar data generation studies, and the logical framework within the applications are not systematically outlined. To bridge this gap, this paper presents a thorough review of 4D mmWave radars in autonomous driving. The principal contributions of this work can be summarized as follows:

\begin{itemize}
    \item{To the best of our knowledge, this is the first publicly available survey focuses on 4D mmWave radars within the context of autonomous driving.}
    \item{Acknowledging the distinctiveness of 4D mmWave radars, this survey outlines its theoretical background, draws a detailed signal processing workflow figure, as the foundation of its application.}
    \item{Given the sparsity and noise of existing 4D mmWave radar point cloud, this paper discusses newly emerged learning-based radar data generation methods that can enhance data quality.}
    \item{This paper delivers an extensive survey of 4D mmWave radar application algorithms in autonomous driving. It systematically presents research on perception and SLAM algorithms of 4D mmWave radars, as well as related datasets, and categorizes them on a timeline in Fig. \ref{Fig:timeline}.}
    \item{By thoroughly tracing relevant research, this paper presents classification framework diagrams for 4D mmWave radar data generation, perception, and SLAM applications. Existing challenges and in-depth future outlook are also illustrated.}
\end{itemize}

The remainder of this paper is organized as shown in Fig. \ref{Fig:Introduction}: Section \ref{Sec:Theory} introduces the foundational theory of 4D mmWave radars, including the signal processing workflow, methods for improving resolution and extrinsic calibration. Section \ref{Sec:Genera} summarizes some learning-based methods for radar data generation. Section \ref{Sec:Percep} reviews 4D mmWave radar perception applications, categorized into different input formats. 4D mmWave radar applications in SLAM are presented in Section \ref{Sec:SLAM}. Moreover, Section \ref{Sec:Datasets} lists available 4D mmWave radar datasets for researchers' convenience. Section \ref{Sec:Future} discusses future trends of 4D mmWave radar in autonomous driving, and Section \ref{Sec:Conclu} draws the conclusion.

\section{Background of 4D mmWave Radars} \label{Sec:Theory}
For researchers dedicated to the field of autonomous driving, fundamental knowledge about 4D mmWave radars may often be undervalued. This section briefly revisits the basic theory of 4D mmWave radars, laying the groundwork for the subsequent discussions.

\subsection{Signal Processing Workflow}\label{subsec:workflow}

\begin{figure*}
    \centering
    \includegraphics[width = \linewidth]{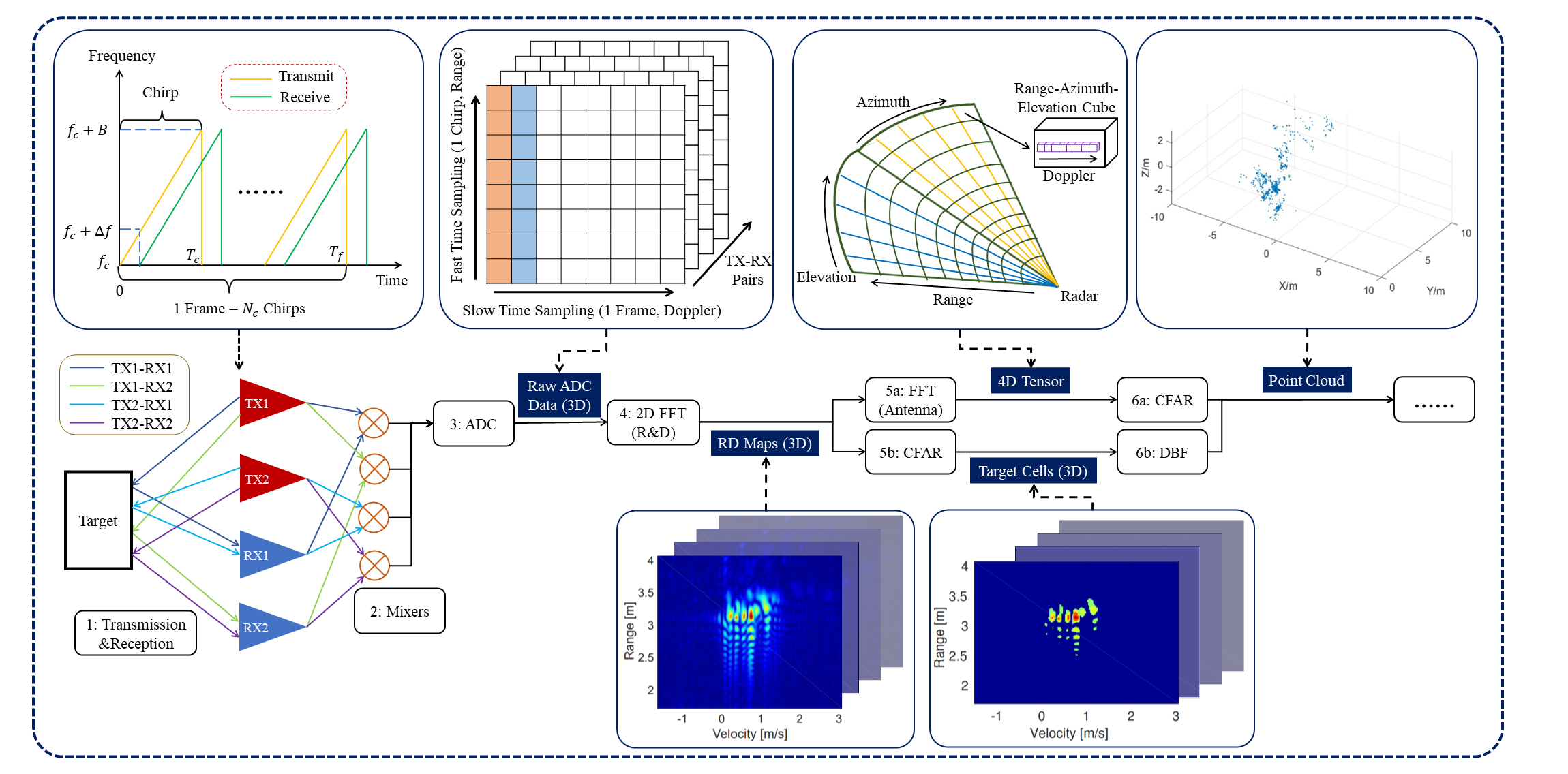}
    \caption{The traditional signal processing workflow and corresponding data formats of 4D mmWave radars \cite{abdulatifMicroDopplerBasedHumanRobot2018} \cite{chengNewAutomotiveRadar2021}}
    \label{Fig:flow}
\end{figure*}

The traditional signal processing workflow and corresponding data formats of 4D mmWave radars are shown in Fig.\ref{Fig:flow}. In step 1, millimeter waves are transmitted from the transmitting (TX) antennas. These waves, upon encountering surrounding targets, are reflected back to receiving (RX) antennas. The waveform employed by the majority of extant 4D mmWave radars is the Frequency Modulated Continuous Wave (FMCW), which  is renowned for its superior resolution capabilities in comparison to alternative waveforms. During each operational cycle (commonly referred to as a 'chirp') of the FMCW radar's TX antennas, the frequency of the emitted signal increases linearly, characterized by an initial frequency $f_c$, a bandwidth $B$, a frequency slope $S$, and a time duration $T_c$. By measuring the frequency of the signal received, the range $r$ of the target can be calculated as follows:
\begin{equation}
    r = \frac{ct}{2}, \quad t = \frac{\Delta f}{S},
\end{equation}
where $t$ denotes the temporal interval between transmission and reception, $c$ represents the light speed, and $\Delta f$ is the discrepancy in frequency between the transmitted and received signals. Concurrently, a single frame of an FMCW radar comprises $N_c$ chirps and spans a temporal duration $T_f$. To avoid interference amongst successive chirps, the transmitted and received signals are considered within an individual chirp. Consequently, the maximum unambiguous range detectable by 4D mmWave radars is restricted by the chirp duration $T_c$. By way of illustration, the AWR1843 from Texas Instruments features a chirp duration of $T_c=0.33\mu s$, accordingly its maximum unambiguous range is 50 meters. Presuming the target's range remains invariant within a single frame, the frequency shift between two successive chirps is employed to deduce the radial relative velocity $v$ of the target, utilizing the Doppler effect, as delineated below:
\begin{equation}
    v = \frac{c\Delta f}{2f_c}, \quad \Delta f = \frac{\Delta \varphi}{2\pi T_c},
\end{equation}
where the first equation is the Doppler effect formula, $\Delta f$ and $\Delta\varphi$ correspond to the frequency and phase shifts, respectively, between the received signals of two successive chirps. It is manifest that the range and Doppler resolutions depend on parameters such as $f_c, T_c, N_c$. For an in-depth exposition of these dependencies, readers are directed to consult the work of Venon et al. \cite{venonMillimeterWaveFMCW2022}. 

The signals of each TX-RX pair are mixed by a mixer at step 2 and subsequently transduced into digital form by an Analog-to-Digital Converter (ADC) at step 3, yielding raw ADC data. It should be noted that within the matrices of raw ADC data depicted in Fig. \ref{Fig:flow}, the coordinate axes represent the sampling timestamps within a chirp and a frame, respectively, while the value of each matrix element corresponds to the intensity of the reflected signal. Sampling within a chirp aims to calculate range information, and is also referred to as fast time sampling. Conversely, sampling within a frame is intended to deduce Doppler information, and is thus termed slow time sampling. Subsequently, at step 4, a two-dimensional Fast Fourier Transformation (FFT) is applied along the range and Doppler dimensions to construct the Range-Doppler (RD) map, the axes of which are range and Doppler velocity. 

\begin{figure}
    \centering
    \includegraphics[width = 0.8\linewidth]{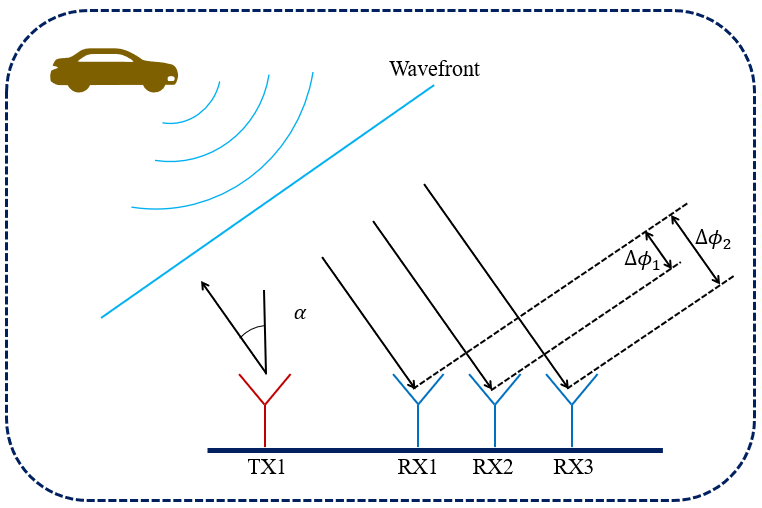}
    \caption{The DOA estimation principle of 4D mmWave radars}
    \label{Fig:DOA}
\end{figure}

However, despite the RD map providing the signal intensities of different ranges and velocities, it does not specify azimuth and elevation angles, rendering the data challenging for humans to understand due to its complex structure. To address this, two prevalent signal processing methodologies are employed to distinguish real objects with high intensity and obtain point clouds. The former is to first conduct a FFT along different TX-RX pairs to deduce the direction-of-arrival (DOA) of the target (step 5a), acquiring a 4D range-azimuth-elevation-Doppler tensor, while for 3D mmWave radars, the result is a 3D range-azimuth-Doppler tensor. Each cell within the 4D tensor corresponds to the intensity of the reflected signal. For DOA estimation, a MIMO antenna design is typically applied in mmWave FMCW radars. As illustrated in Fig. \ref{Fig:DOA}, the $n$ TX antennas and $m$ RX antennas form $n \times m$ virtual TX-RX pairs. To ensure signal separation, different TX antennas should transmit orthogonal signals. By analyzing the phase shift between different TX-RX pairs, distance differences between different pairs to the same target can be calculated. Furthermore, by considering the positional arrangement of the TX and RX antennas, the DOA of the target can be ascertained. At step 6a, the Constant False Alarm Rate (CFAR) algorithm \cite{rohlingRadarCFARThresholding1983} is typically implemented in the four dimensions to filter the tensor based on the intensity of each cell, thereby obtaining real targets in the format of point cloud for subsequent applications \cite{kramerColoRadarDirect3D2022}. The CFAR algorithm sets dynamic intensity thresholds by comparing the intensity of each cell with its neighboring cells to realize a constant false alarm rate effect.


In contrast, the alternative signal processing workflow initially filters RD maps to generate target cells using also a CFAR-type algorithm (step 5b), then digital beamforming (DBF) is employed in step 6b to recover angular information and generate point clouds \cite{chengNewAutomotiveRadar2021}.

\subsection{Methods to Lift from 3D to 4D}
As previously discussed, the most crucial ability of 4D mmWave radars lies in their ability to measure the elevation dimension, which enriches the data from three-dimensional to four-dimensional space. The methodologies to achieve this enhancement can be categorized into hardware-based and software-based approaches, as detailed below:

\subsubsection{Hardware}
At the hardware level, there are two principal strategies to improve elevation resolution. The first is to increase the number of TX-RX pairs by simply cascading multiple standard mmWave radar chips \cite{ochScalable77GHz2018} or integrating more antennas onto a single chip \cite{charlesFullyIntegrated782021}. The second strategy aims to the effective aperture of the antennas by techniques such as meta-material \cite{jiangWidebandMIMODirectional2020}.

\subsubsection{Software}
By virtually realizing hardware improvement or optimizing signal processing algorithms along the processing workflow, radar resolution can also be improved at the software level. Inspired by the synthetic aperture radar (SAR) technology, angular resolution can be increased by virtually expanding the aperture of antennas through software design \cite{wuGeneralizedThreeDimensionalImaging2019}. Furthermore, innovative learning-based algorithms have the potential to replace traditional signal processing algorithms, such as FFT and CFAR \cite{choGuidedGenerativeAdversarial2021} \cite{brodeskiDeepRadarDetector2019} thus facilitating a super-resolution effect. 

\subsection{Extrinsic Calibration}

Given the relative sparsity and noise of radar point clouds, and the non-intuitive nature of spectrum data, it is a significant challenge to calibrate radars with other sensors. While the enhanced resolution of 4D mmWave radars somewhat mitigates this issue, there remains a dearth of robust online calibration methods.


Following traditional calibration methods of 3D mmWave radars, corner reflectors are commonly employed to improve calibration accuracy. By carefully placing several corner reflectors and analyzing the sensing results of the 4D mmWave radar in conjunction with LiDAR and camera data, the extrinsic parameters can be calibrated \cite{zhengTJ4DRadSet4DRadar2022}. In a departure from the conventional approach of calibrating each sensor pair sequentially, Domhof et al. calibrate all sensors simultaneously with respect to the body of mobile robot, achieving a median rotation error of a mere 0.02 $^{\circ}$ \cite{domhofJointExtrinsicCalibration2021}. By leveraging the Random Sample and Consensus (RANSAC) and Levenberg-Marquardt non-linear optimization, \cite{cheng3DRadarCamera2023} accomplishes radar-camera calibration with only one single corner reflector, obviating the requirement of a specially designed calibration environment.

However, the practicability of corner reflectors in real-world scenarios is limited. Recent research has proposed calibration methods for 4D mmWave radars that avoid the need for specially placed corner reflectors, instead utilizing radar motion measurement to conduct online calibration for radars \cite{baoMotionBasedOnline2020} or radar-camera pairs \cite{wiseContinuousTimeApproach3D2021}. While these methods offer convenience, their efficacy under extreme weather conditions remains to be validated.

In light of the similar data structures of 4D mmWave radars and LiDARs, modifying conventional LiDAR-to-camera calibration methods \cite{dhallLiDARcameraCalibrationUsing2017} \cite{pusztaiAccurateCalibrationLiDARcamera2017} is a promising avenue. Nevertheless, to address online joint calibration in extreme weather conditions, especially in the situation where LiDAR and camera have lousy performance, the potential of learning-based methods \cite{schollerTargetlessRotationalAutoCalibration2019} warrants further exploration.

\section{Learning-based Radar Data Generation} \label{Sec:Genera}

As discussed in Section \ref{Sec:Theory}, the initial data from 4D mmWave radars comprise spectral signals heavily masked by noise, necessitating filtering algorithms like CFAR to generate usable point clouds. However, such traditional handcrafted methods have inherent limitations. They often struggle to adapt to the complexity of real-world targets, which can vary significantly in shape and extend across multiple resolution cells. This mismatch can induce masking effects within CFAR-type algorithms, consequently reducing the resolution of point clouds and resulting in significant information loss.

\begin{figure}[htbp]
    \begin{minipage}[t]{0.45\linewidth}
        \centering
        \includegraphics[width=\textwidth]{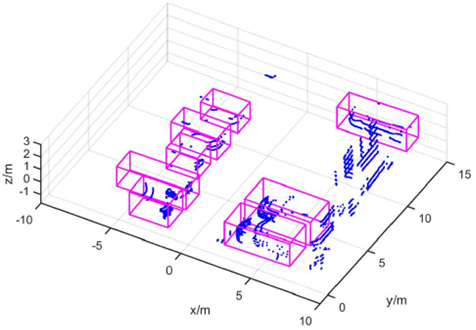}
        {\scriptsize (a) LiDAR point cloud}
    \end{minipage}
    \hfill
    \begin{minipage}[t]{0.45\linewidth}
        \centering
        \includegraphics[width=\textwidth]{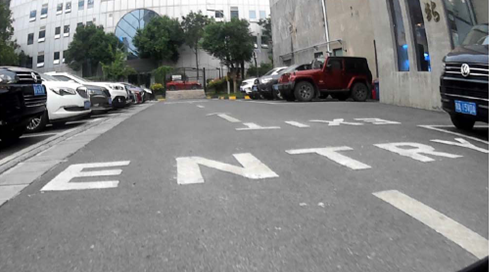}
        {\scriptsize (b) RGB image}
    \end{minipage}
    
    \vspace{2ex}
    
    \begin{minipage}[t]{0.45\linewidth}
        \centering
        \includegraphics[width=\textwidth]{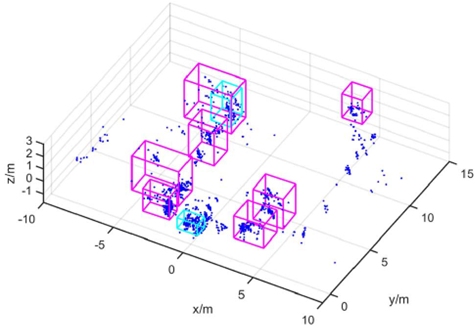}
        {\scriptsize (c) OS-CFAR point cloud}
    \end{minipage}
    \hfill
    \begin{minipage}[t]{0.45\linewidth}
        \centering
        \includegraphics[width=\textwidth]{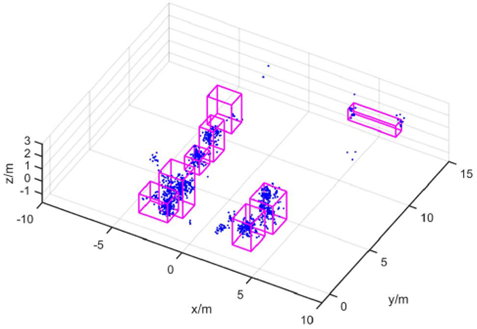}
        {\scriptsize (d) NN-Generated point cloud}
    \end{minipage}
    
\caption{Object detection results from LiDAR point clouds and radar point clouds generated by the OS-CFAR detector and learning-based detector\cite{chengNovelRadarPoint2022}.}
\label{Fig:generation-results}
\end{figure}

To overcome those limitations, this section introduces learning-based techniques for radar data generation. By leveraging the capabilities of deep learning, it is possible to develop more adaptive and robust algorithms that can improve the fidelity of radar imaging. As illustrated in Fig. \ref{Fig:generation-results}, radar point clouds generated using a learning-based detector are denser and contain less noise compared to those produced by the traditional OS-CFAR detector.

\begin{figure}[htbp]
    \centering
    \includegraphics[width = 0.9\linewidth]{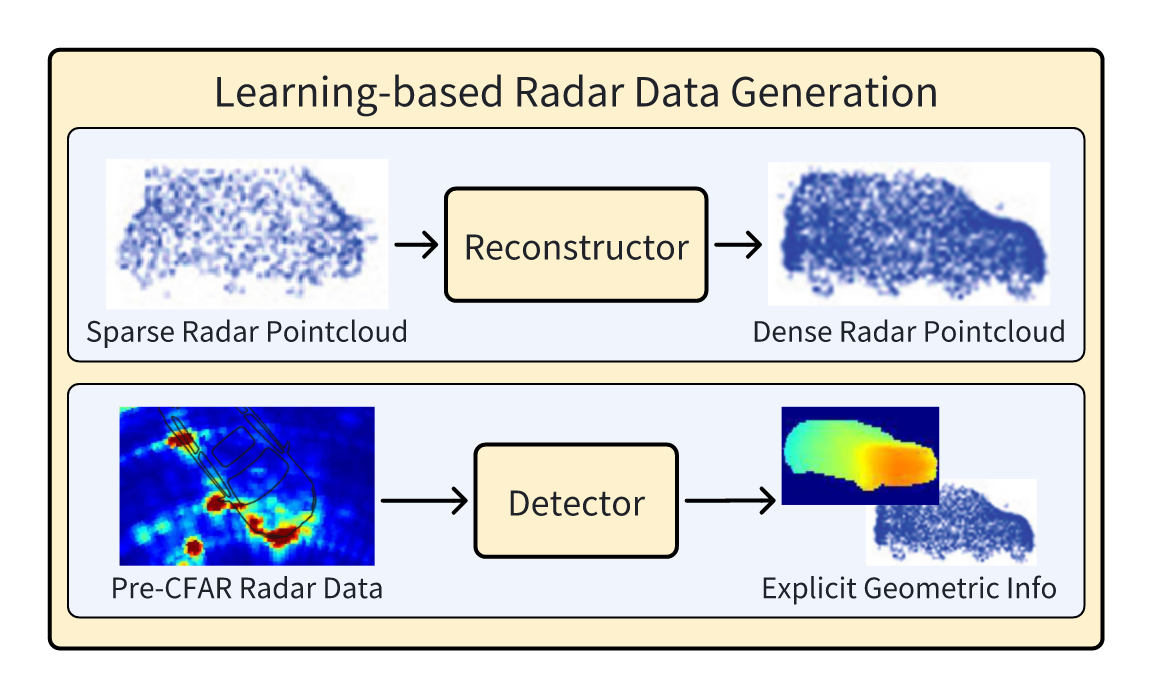}
    \caption{Two dominant pipelines in learning-based radar data generation.}
    \label{Fig:Improvement}
\end{figure}

In the current landscape of mmWave radar technology, two primary learning-based pipelines have been developed: "Reconstructor" and "Detector," as shown in Fig. \ref{Fig:Improvement}. Reconstructor techniques focus on refining radar point clouds by enhancing their density and resolution. In contrast, Detector methods bypass the preliminary CFAR filtering stage and process radar frequency data directly, thus preventing the information loss typically associated with traditional filtering methods. Subsequent sections will provide a detailed comparison of these approaches and discuss the persistent challenges in this area.

\subsection{Reconstructor}

Reconstructor methods focus on improving the resolution and detail of previously acquired radar point clouds. This approach is dedicated to the post-processing enhancement of data fidelity, thereby increasing the usefulness of the radar imagery.

\begin{figure}[htbp]
\centerline{\includegraphics[width = \linewidth]{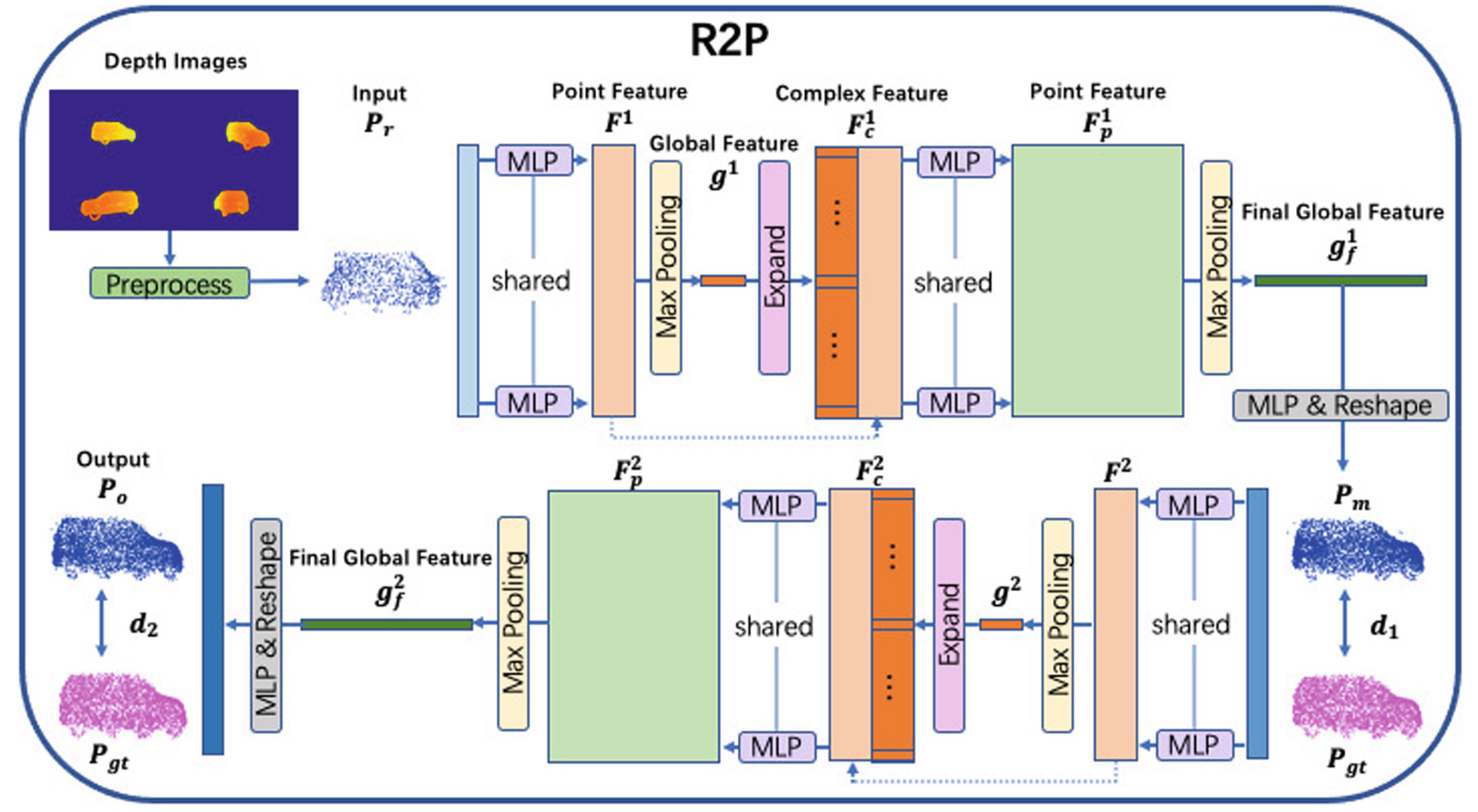}}
\caption{R2P\cite{sunR2PDeepLearning2022} network architecture}
\label{Fig:R2P}
\end{figure}

Much of the inspiration for these methods comes from  the reconstruction of LiDAR point clouds. Notably, the PointNet structure \cite{charlesPointNetDeepLearning2021} used in \cite{yuanPCNPointCompletion2018} has influenced subsequent studies by Sun et al. \cite{sun3DRIMR3DReconstruction2021} \cite{sunR2PDeepLearning2022} \cite{sun3DReconstructionMultiple2022}. Although these studies require data from multiple viewpoints, which may constrain their immediate integration into autonomous driving systems, the underlying principles of their model architectures offer valuable insights. For example, they use a conditional Generative Adversarial Network (GAN) to train generator and discriminator networks concurrently, as detailed in \cite{sun3DRIMR3DReconstruction2021}. Moreover, the innovative two-stage point cloud generation process, which incorporates a loss function that synergistically combines Chamfer Distance (CD) and Earth Mover's Distance (EMD) metrics, is described in \cite{sunR2PDeepLearning2022}. Sun et al.'s methods have shown significant improvements over existing techniques such as PointNet \cite{charlesPointNetDeepLearning2021}, PointNet++ \cite{qiPointNetDeepHierarchical2017}, and PCN \cite{yuanPCNPointCompletion2018}, particularly with coarse and sparse input point clouds. The robustness of these methods underscores their potential to enhance the accuracy and reliability of point cloud reconstruction, even with suboptimal radar data.

\subsection{Detector}

Meanwhile, Detector approaches leverage neural networks to engage directly with raw radar data such as RD maps or 4D tensors, which circumvents conventional techniques such as CFAR or DBF, potentially leading to more efficient and robust detection capabilities in real-time applications.

Brodeski et al. \cite{brodeskiDeepRadarDetector2019} pioneer such frameworks and apply CNN-based image segmentation networks to RD maps for the detection and localization of multiple objects. Confronted with the scarcity of well-annotated RD map datasets, they devise a strategy to extract labeled radar data from the calibration process conducted within an anechoic chamber. Experiments of the DRD network demonstrate its capability to function in real-time, with inference times recorded at approximately 20ms. Notably, the DRD network has been shown to surpass classic methods in terms of detection accuracy and robustness. Though this work does not include real-world radar data with all the impairments that come along, the findings from this study unequivocally illustrate the considerable promise of neural network applications to radar complex data.

However, accurately labeling radar frequency data remains a formidable challenge. This is primarily due to disparities between data collected in the controlled environment of anechoic chambers and that obtained under real-world driving scenarios. The latter presents greater complexity with factors such as multi-path reflections, interference, attenuation, etc. 

\begin{figure}[htbp]
\centerline{\includegraphics[width = \linewidth]{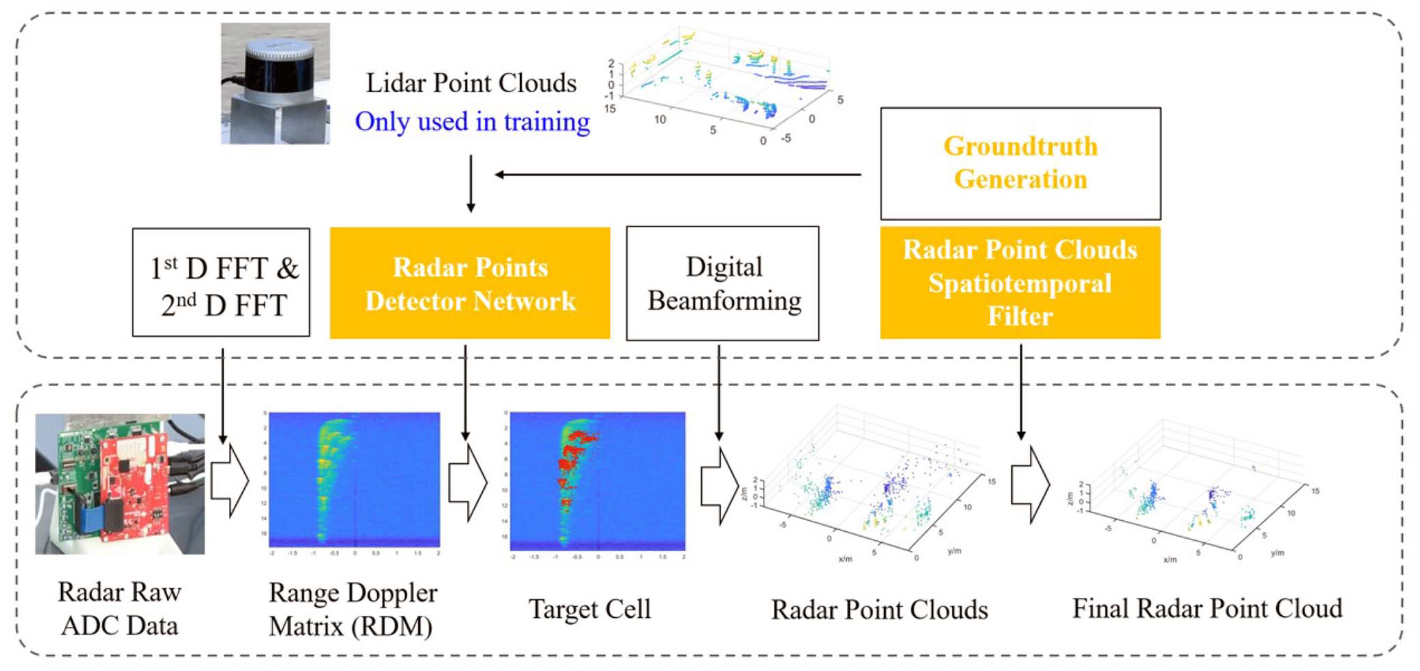}}
\caption{Overview of the radar signal processing chain with \cite{chengNovelRadarPoint2022}}
\label{Fig:RPC_generation_zhang}
\end{figure}

To address this challenge, Cheng et al. \cite{chengNewAutomotiveRadar2021,chengNovelRadarPoint2022} use LiDAR point clouds as the supervision and successively design network architectures inspired by U-Net\cite{ronnebergerUNetConvolutionalNetworks2015} and Generative Adversarial Networks (GAN)\cite{lucSemanticSegmentationUsing2016}. In complex roadway scenes, the generated 4D mmWave radar point clouds by \cite{chengNovelRadarPoint2022} not only demonstrate a reduction in clutter but also provide denser point clouds of real targets compared to the classical CFAR detectors. Additional comparisons on the performance of perception and localization tasks between the generated point cloud and traditional point cloud further prove the improvement of data quality.

\subsection{Challenge}


The development of Learning-based Radar Data Generation methods, particularly for 4D mmWave radar, is hindered by the scarcity of large, public datasets and benchmarks. As highlighted in recent studies \cite{madaniRadatronAccurateDetection2022,orrHighresolutionRadarRoad2021}, the diversity of mmWave radar hardwares models complicates the standardization of pre-CFAR data andchallenges the creation of a comprehensive public dataset and benchmark. Furthermore, pre-CFAR data is far less intuitive than point cloud data, rendering manual annotation both laborious and prone to errors, thus complicating the production of high-quality supervised data for neural networks. 

Another avenue is the adoption of learning-based approaches for generating synthetic automotive radar scenes and data\cite{chipengoHighFidelityPhysicsBased2023,tanLearningbased4DMillimeter2023}. However, as such methods are based on simulation, it is limited by potential inaccuracies in the sensor and world model. 

Moreover, the management of pre-CFAR data is associated with considerable computational and memory demands. While current methods employ lower-resolution radars to achieve real-time performance, accommodating higher resolution radars—which are essential for large outdoor scenes—requires further optimization of algorithmic efficiency.

\begin{figure*}[htbp]
    \centering
    \includegraphics[width = \linewidth]{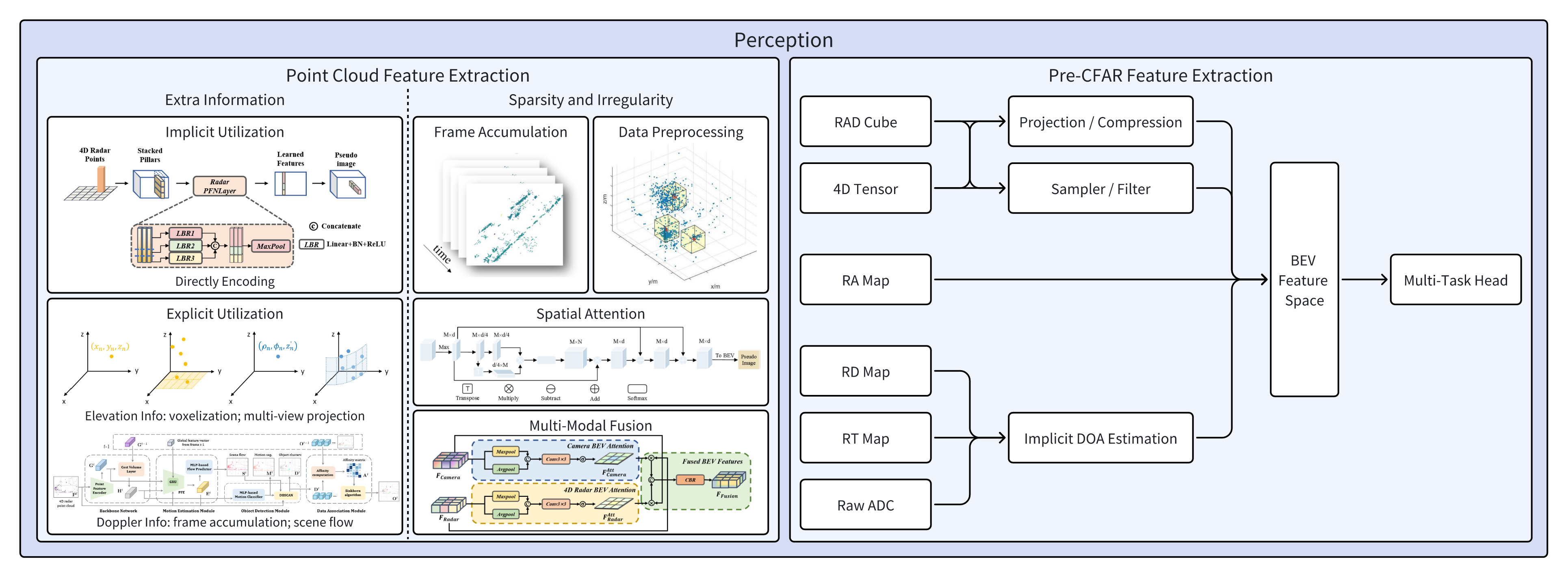}
    \caption{Feature extraction methods for 4D mmWave radars in autonomous perception. Example images are from \cite{zhengRCFusionFusing4D2023,yanMVFANMultiViewFeature2023,tan3DObjectDetection2023,liuSMURFSpatialMultiRepresentation2023,xuRPFANet4DRaDAR2021}.}
    \label{Fig:Perception}
\end{figure*}

\section{Perception Applications} \label{Sec:Percep}



Currently, the point cloud density of 4D mmWave radars has already attained a level comparable to that of low-beam LiDAR, with the added advantages of exhibiting superior robustness under low visibility and adverse weather conditions. Therefore, researchers have been attempting to transfer LiDAR point cloud processing models to 4D mmWave radars in various tasks, including target detection\cite{cui3DDetectionTracking2021,liuSMURFSpatialMultiRepresentation2023,wangMultiModalMultiScaleFusion2023}, trajectory tracking\cite{panMovingObjectDetection2023,tanTrackingMultipleStatic2023}, and scene flow prediction\cite{dingSelfSupervisedSceneFlow2022,dingHiddenGems4D2023}, among others. Furthermore, as described in Section \ref{Sec:Genera}, pre-CFAR radar data encompasses a wealth of information, promoting some researchers to engage directly with RD maps or 4D tensors, bypassing point cloud generation tasks. Existing 4D mmWave radar point cloud and pre-CFAR data feature extraction methods in autonomous driving perception are summarized in Fig. \ref{Fig:Perception}, which will be detailed in this section.

In Section \ref{Sec:Percept-PC}, we review and analyze perception models for radar point clouds (RPC), which are primarily enhancements of LiDAR-based methodologies. The 4D mmWave radar branch of some fusion methods are also included.
Section \ref{Sec:Percept-preCFAR} investigates the methods utilizing pre-CFAR radar data, including the range-frequency map, range-azimuth map, range-azimuth-Doppler cube, and 4D tensors. 
The integration of 4D mmWave radars into multi-modal fusion systems, as well as the effectiveness of such methods are present in Section \ref{Sec:Percept-fusion}.
Finally in Section \ref{Sec:Percept-challenge}, we discuss current challenges of this field.

\begin{sidewaystable*}[htp]
\caption{\MakeUppercase{Summary of 4D mmWave Radar Perception Methods}} 
\label{Tab:4D-Radar perception methods}
\centering
\small
\begin{threeparttable}
\begin{tabular}{ccccccccccc}
\multicolumn{11}{c}{\MakeUppercase{a. with radar pointcloud}}\\
\toprule
\multirow{2}{*}{Task}                         & \multirow{2}{*}{Methods}                                            & \multirow{2}{*}{Year} & \multirow{2}{*}{Modalities} & \multicolumn{3}{c}{Astyx 3D mAP(\%)}                                                                                      & \multicolumn{2}{c}{VoD\cite{palffyMultiClassRoadUser2022}   mAP(\%)} & \multicolumn{2}{c}{TJ4D mAP(\%)}                                \\
                                              &                                                                     &                       &                             & Easy                                                                        & Moderate             & Hard                 & Entire                            & Driving                          & 3D                             & BEV                            \\ 
\midrule
3D Object Detection                           & PointPillars\cite{langPointPillarsFastEncoders2019}\ddag            & 2019                  & RPC                           & 30.14                                                                       & 24.06                & 21.91                & 38.09                             & 62.58                            & 28.31                          & 36.23                          \\
3D Object Detection                           & CenterPoint\cite{yinCenterbased3DObject2021}\ddag                   & 2021                  & RPC                           & -                                                                           & -                    & -                    & 45.42\dag                         & 65.06\dag                        & 29.07                          & 36.18                          \\
3D Object Detection                           & PillarNeXt\cite{liPillarNeXtRethinkingNetwork2023}\ddag             & 2023                  & RPC                           & -                                                                           & -                    & -                    & 42.23\dag                         & 63.61\dag                        & 29.20                          & 35.71                          \\ 
\midrule
3D Object Detection                           & RPFA-Net\cite{xuRPFANet4DRaDAR2021}                                 & 2021                  & RPC                           & 38.85                                                                       & 32.19                & 30.57                & 38.75                             & 62.44                            & 29.91                          & 38.94                          \\
3D Object Detection                           & MVFAN\cite{yanMVFANMultiViewFeature2023}                            & 2023                  & RPC                  & \underline{\textbf{45.60}}                                                        & \underline{\textbf{39.52}} & \underline{\textbf{38.53}} & \underline{\textbf{39.42}}              & \underline{\textbf{64.38}}             & -                              & -                              \\
3D Object Detection                           & RadarPillarNet\cite{zhengRCFusionFusing4D2023}                      & 2023                  & RPC                           & -                                                                           & -                    & -                    & 46.01\dag                         & 65.86\dag                        & 30.37                          & 39.24                          \\
3D Object Detection                           & LXL-R\cite{xiongLXLLiDARExcluded2023}                               & 2023                  & RPC                           & -                                                                           & -                    & -                    & 46.84\dag                         & 68.51\dag                        & 30.79                          & 38.42                          \\
3D Object Detection                           & SMIFormer\cite{shiSMIFormerLearningSpatial2023}                     & 2023                  & RPC                           & -                                                                           & -                    & -                    & \textbf{48.77\dag}                & \underline{\textbf{71.13\dag}}         & -                              & -                              \\
3D Object Detection                           & SMURF\cite{liuSMURFSpatialMultiRepresentation2023}                  & 2023                  & RPC                           & -                                                                           & -                    & \textbf{-}           & \underline{\textbf{50.97\dag}}          & \textbf{69.72\dag}               & \underline{\textbf{32.99}}           & \underline{\textbf{40.98}}           \\
3D Object Detection                           & RadarMFNet\cite{tan3DObjectDetection2023}                           & 2023                  & RPC                           & -                                                                           & -                    & -                    & -                                 & \textbf{-}                       & \underline{\textbf{42.61\dag}}       & \underline{\textbf{49.07\dag}}       \\ 
\midrule
3D Object   Detection                         & FUTR3D\cite{chenFUTR3DUnifiedSensor2023}\ddag                       & 2023                  & C\&RPC                        & -                                                                           & -                    & -                    & 49.03\dag                         & 69.32\dag                        & 32.42                          & 37.51                          \\
3D Object Detection                           & BEVFusion\cite{liuBEVFusionMultiTaskMultiSensor2023}\ddag           & 2023                  & C\&RPC                        & -                                                                           & -                    & -                    & 49.25\dag                         & 68.52\dag                        & 32.71                          & 41.12                          \\ 
\midrule
3D Object Detection                           & 3DRC\cite{meyerDeepLearningBased2019}                               & 2019                  & C\&RPC                        & 61.00                                                                       & 48.00                & 45.00                & -                                 & -                                & -                              & -                              \\
3D Object Detection                           & Cui et al.\cite{cui3DDetectionTracking2021}                         & 2021                  & C\&RPC              & \underline{\textbf{69.50}}                                                        & \underline{\textbf{50.05}} & \underline{\textbf{49.13}} & -                                 & -                                & -                              & -                              \\
3D Object Detection                           & RCFusion\cite{zhengRCFusionFusing4D2023}                            & 2023                  & C\&RPC                        & -                                                                           & -                    & -                    & 49.65                             & 69.23                            & 33.85                          & 39.76                          \\
3D Object Detection                           & LXL\cite{xiongLXLLiDARExcluded2023}                                 & 2023                  & C\&RPC                        & -                                                                           & -                    & \textbf{-}           & \underline{\textbf{56.31}}              & \underline{\textbf{72.93}}             & \underline{\textbf{36.32}}           & \underline{\textbf{41.20}}           \\
3D Object Detection                           & InterFusion\cite{wangInterFusionInteractionbased4D2022}             & 2022                  & L\&RPC                        & 57.07                                                                       & 47.76                & 45.05                & -                                 & -                                & -                              & -                              \\
3D Object Detection                           & $M^2$-Fusion\cite{wangMultiModalMultiScaleFusion2023}               & 2023                  & L\&RPC               & \underline{\textbf{61.33}}                                                        & \underline{\textbf{49.85}} & \underline{\textbf{49.12}} & -                                 & -                                & -                              & -                              \\ 
\midrule
 & & & & & & & & & \\
\multicolumn{11}{c}{\MakeUppercase{b. with pre-CFAR data}}\\
\midrule
\multirow{2}{*}{Task}                         & \multirow{2}{*}{Methods}                                            & \multirow{2}{*}{Year} & \multirow{2}{*}{Modalities} & \multirow{2}{*}{\begin{tabular}[c]{@{}c@{}}Radar\\      Input\end{tabular}} & \multicolumn{4}{c}{RADIal\cite{rebutRawHighDefinitionRadar2022}(\%)}                                               & \multicolumn{2}{c}{K-Radar\cite{paekKRadar4DRadar2022} mAP(\%)} \\
                                              &                                                                     &                       &                             &                                                                             & Seg. mIoU            & Det. AP              & Det. AR                           & 3D Det. mAP                      & 3D                             & BEV                            \\
\midrule
Freespace Segmentation; 2D   Object Detection & T-FFTRadNet\cite{girouxTFFTRadNetObjectDetection2023}               & 2023                  & R                           & ADC                                                                         & 79.60                & 88.20                & 86.70                             & -                                & -                              & -                              \\
Freespace Segmentation; 2D   Object Detection & T-FFTRadNet\cite{girouxTFFTRadNetObjectDetection2023}               & 2023                  & R                           & RD                                                                          & 80.20                & 89.60                & 89.50                             & -                                & -                              & -                              \\
Freespace Segmentation; 2D   Object Detection & ADCNet\cite{yangADCNetLearningRaw2023}                              & 2023                  & R                           & ADC                                                                         & 78.59                & 95.00                & 89.00                             & -                                & -                              & -                              \\
Freespace Segmentation; 2D   Object Detection & FFTDASHNet\cite{bootEfficientDASHAutomatedRadar2023}                & 2023                  & R                           & RD                                                                          & \underline{\textbf{85.58}} & 96.53                & \underline{\textbf{98.51}}              & -                                & -                              & -                              \\
Freespace Segmentation; 2D   Object Detection & FFTRadNet\cite{rebutRawHighDefinitionRadar2022}                     & 2022                  & R                           & RD                                                                          & 73.98                & 96.84                & 82.18                             & -                                & -                              & -                              \\
Freespace Segmentation; 2D   Object Detection & TransRadar\cite{dalbahTransRadarAdaptiveDirectionalTransformer2024} & 2024                  & R                           & RAD                                                                         & 81.10                & \underline{\textbf{97.30}} & 98.40                             & -                                & -                              & -                              \\
Freespace Segmentation; 2D   Object Detection & CMS\cite{jinCrossModalSupervisionBasedMultitask2023}                & 2023                  & C\&R                        & RD                                                                          & 80.40                & 96.90                & 83.49                             & -                                & -                              & -                              \\
2D \& 3D Object Detection                     & EchoFusion\cite{liuEchoesPointsUnleashing2023}                      & 2023                  & C\&R                        & RT                                                                          & -                    & 96.95                & 93.43                             & \underline{\textbf{39.81}}             & \underline{\textbf{68.35*}}           & \underline{\textbf{69.95*}}           \\
3D Object Detection                           & RTN\cite{paekKRadar4DRadar2022}                                     & 2022                  & R                           & 4DRT                                                                        & -                    & -                    & -                                 & -                                & 40.12                          & 50.67                          \\
3D Object Detection                           & RTNH\cite{paekKRadar4DRadar2022}                                    & 2022                  & R                           & 4DRT                                                                        & -                    & -                    & -                                 & -                                & 47.44                          & 58.39                          \\
3D Object Detection                           & E-RTNH\cite{paekEnhancedKRadarOptimal2023}                          & 2023                  & R                           & 4DRT                                                                        & -                    & -                    & -                                 & -                                & 47.90                          & 59.40                         \\ 
\bottomrule
\end{tabular}
\begin{tablenotes}
\item Abbreviation about Modalities and Radar Input: R(Radar), C(Camera), L(Lidar), RPC(Radar Point Cloud), ADC(raw radar data after Analog-to-Digital Converter), RD(Range-Doppler map), RAD(Range-Azimuth-Doppler cube), RT(Range-Time map), 4DRT(Range-Azimuth-Elevation-Doppler Tensor)
\item \dag \quad indicates data derived through multi-frame accumulation. Specifically, the methodologies referenced in relation to the VoD dataset employ detection points from 5 scans of radar data, whereas the RadarMFNet\cite{tan3DObjectDetection2023} approach, as applied to the TJ4DRadSet, utilizes data from 4 consecutive frames.
\item \ddag \quad denotes strategies originally conceptualized for LiDAR point clouds, which have been subsequently adapted and retrained utilizing radar datasets to serve as baseline comparisons. The comparative outcomes are directly inherated from \cite{xuRPFANet4DRaDAR2021,yanMVFANMultiViewFeature2023,zhengRCFusionFusing4D2023,liuSMURFSpatialMultiRepresentation2023,xiongLXLLiDARExcluded2023}.
\item * \ signifies data extracted from a subset comprising 20 sequences, which is part of the K-Radar dataset encompassing a total of 58 sequences.
\end{tablenotes}
\end{threeparttable}
\end{sidewaystable*}

\subsection{Point Cloud Feature Extraction}\label{Sec:Percept-PC}

Given the analogous nature of their data formats, it is clear that a significant number of RPC methodologies originate from LiDAR-based techniques. Despite this similarity, this transposition requires careful consideration of the inherent constraints associated with radar systems. These limitations include low-resolution representations, data sparsity, and inherent uncertainty within the data. Conversely, radar systems outperform in areas such as superior range resolution, velocity measurement capabilities, and early target detection. Consequently, these unique characteristics necessitate the development of specifically tailored network designs.


A comprehensive and succinct overview of recent advances in this field is presented in Table \ref{Tab:4D-Radar perception methods}. These investigations have masterfully exploited the distinctive attributes of 4D mmWave radar point clouds, encompassing elements like elevation, Doppler data, and Radar Cross Section (RCS) intensity. Moreover, they have ingeniously formulated strategies to address the inherent sparsity and irregular distribution of these data points, thereby advancing the field significantly.

\subsubsection{Distinctive Information}


4D mmWave radars provide a full three-dimensional view by measuring range, azimuth, and elevation of targets. Additionally, mmWave radars can measure the velocity of objects directly through the Doppler effect, a feature that distinguishes them from LiDAR systems. This combination of enhanced spatial data and velocity information makes 4D mmWave radars particularly valuable for Autonomous Driving tasks and presents unique opportunities for further studies.


Most studies\cite{meyerDeepLearningBased2019,xuRPFANet4DRaDAR2021,tan3DObjectDetection2023,palffyMultiClassRoadUser2022,wangInterFusionInteractionbased4D2022} have opted to reference the implicit structures like SECOND\cite{yanSECONDSparselyEmbedded2018} and PointPillars\cite{langPointPillarsFastEncoders2019}. These studies encode extra radar attributes directly, comparable to the conventional spatial coordinates $x, y, z$ in point clouds. Palffy et al.\cite{palffyMultiClassRoadUser2022} demonstrate that the addition of elevation data, Doppler information, and RCS information respectively increase the 3D mean Average Precision (mAP) with 6.1\%, 8.9\% and 1.4\%. However, the result of the proposed method (47.0\% mAP) is still far inferior to the LiDAR detector on 64-beam LiDAR (62.1\% mAP), indicating that there is still room for improvement in the optimization of 4D mmWave radar-based detection methods.Nevertheless, Zheng et al.\cite{zhengRCFusionFusing4D2023} introduce a subtle yet impactful modification, by proposing the Radar PillarNet backbone, colloquially termed RPNet. This structure employs three separate linear layers, each with unshared weights, to extract spatial position, velocity, and intensity features, respectively. Subsequently, a BEV pseudo image is generated. Ablation studies have demonstrated that RPNet enhances the 3D mAP by 4.26\%.

\begin{figure}[htbp]
\centerline{\includegraphics[width = \linewidth]{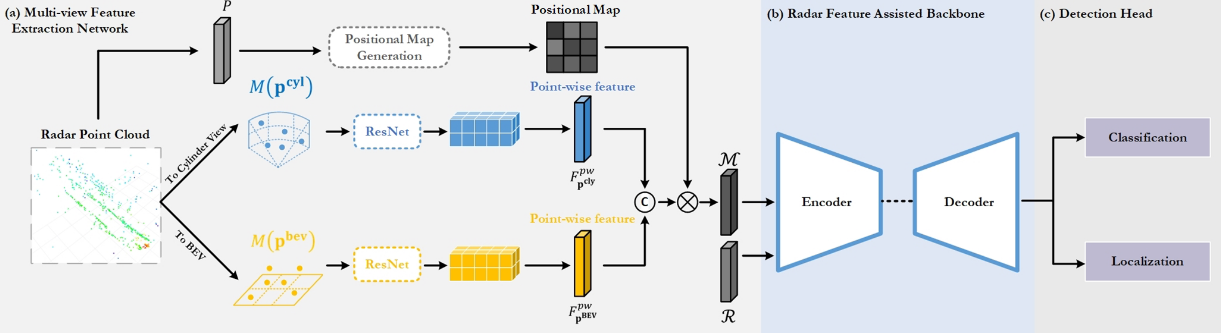}}
\caption{The flowchart of multi-view feature extraction \cite{yanMVFANMultiViewFeature2023}}
\label{Fig:MVFAN}
\end{figure}


Furthermore, to explicitly utilize the elevation information, Cui et al.\cite{cui3DDetectionTracking2021}, Yan et al.\cite{yanMVFANMultiViewFeature2023} and Shi et al.\cite{shiSMIFormerLearningSpatial2023} have each explored extracting point cloud features from multiple viewpoints. In \cite{cui3DDetectionTracking2021}, radar point clouds are processed into Front View (FV) and BEV perspectives. Features extracted from each view are subsequently fused with features derived from the camera branch. MVFAN\cite{yanMVFANMultiViewFeature2023} employs both BEV pillar and cylinder pillar methods to extract point cloud features. Conversely, SMIFormer\cite{shiSMIFormerLearningSpatial2023} transforms point clouds into voxel features, which are then projected onto three distinct planes: FV, Side View (SV), and BEV. Following this, features are aggregated using intra-view self-attention and inter-view cross-attention mechanisms. Notably, this methodology is further refined by employing a sparse dimension compression technique, significantly reducing the memory and computational demands involved in converting 3D voxel features into 2D features.


Addressing the explicit utilization of Doppler information, Tan et al. \cite{tan3DObjectDetection2023} delineate a widely-adopted yet efficacious technique. Recognizing that mmWave radars intrinsically measure the radial relative velocity of detected objects, and the motion of the vehicle itself results in different coordinate systems for multi-frame point clouds, they first calculate the vehicle's velocity, followed by compensating and obtaining each point's true velocity relative to the ground, which is often referred to as 'absolute velocity'. Moreover, the integration of Doppler information facilitates a more convenient and accurate accumulation of historical frame point clouds, thereby enhancing point cloud density. Yan et al. \cite{yanMVFANMultiViewFeature2023} further propose the Radar Feature Assisted Backbone. In this design, each point's absolute and relative velocities, along with its reflectivity, are integrated into position embeddings. These embeddings are then multiplied with the self-attention reweighting map of point-wise features, thereby enhancing the exchange of information at the feature vector level in a trainable fashion. On the other hand, Pan et al. \cite{panMovingObjectDetection2023} introduce a 'detection by tracking' strategy. This approach leverages velocity characteristics to achieve point-level motion segmentation and scene flow estimation. Subsequently, employing the classical DBSCAN clustering method suffices to surpass the tracking accuracy of established techniques like centerpoint\cite{yinCenterbased3DObject2021} and AB3DMOT\cite{weng3DMultiObjectTracking2020}.


\subsubsection{sparsity and Irregularity}


Another significant challenge in processing 4D mmWave radar point clouds is the inherent sparsity and irregular distribution. Considering the physical size constraints on the aperture of vehicular radars and the omnipresence of electromagnetic interference and multipath reflections in traffic environments, the resolution of 4D mmWave radar point clouds is considerably inferior to that of LiDAR, often resulting in a higher prevalence of clutter and noise. For instance, studies have noted that point clouds in the Astyx dataset struggle to articulate detailed features, which complicates the assessment of the orientation of detected objects \cite{xuRPFANet4DRaDAR2021}. Moreover, a considerable number of points are found to be distributed below the ground plane \cite{wangMultiModalMultiScaleFusion2023}, adversely affecting detection accuracy.

To mitigate these challenges, several improvement strategies have been proposed and employed. Common methods include the accumulation of multiple frame point clouds \cite{tan3DObjectDetection2023, panMovingObjectDetection2023}, preprocessing and filtering of point clouds \cite{xuRPFANet4DRaDAR2021,wangMultiModalMultiScaleFusion2023,liuSMURFSpatialMultiRepresentation2023}, employing spatial attention mechanisms to extract contextual information for feature enhancement \cite{xuRPFANet4DRaDAR2021,shiSMIFormerLearningSpatial2023,zhengRCFusionFusing4D2023,yanMVFANMultiViewFeature2023}, and integrating information from different sensor modalities \cite{wangInterFusionInteractionbased4D2022,zhengRCFusionFusing4D2023,wangMultiModalMultiScaleFusion2023,xiongLXLLiDARExcluded2023}.


To accumulate point clouds across multiple consecutive frames, as previously mentioned, the Doppler information plays a pivotal role. This can be achieved through ego-velocity estimation and motion compensation \cite{tan3DObjectDetection2023}, or by motion segmentation and scene flow estimation \cite{panMovingObjectDetection2023}, resulting in precision that surpasses mere simple stacking of point clouds.


Targeting at the preprocessing phase, InterFusion\cite{wangInterFusionInteractionbased4D2022} and M2Fusion\cite{wangMultiModalMultiScaleFusion2023} utilize a Gaussian normal distribution to assess whether the vertical angle of point falls within a normal range, based on the Shapiro-Wilk (S-W) test\cite{puriAugmentingShapiroWilkTest1976}. This approach effectively filters out a substantial number of noise points that are below the ground plane in the Astyx dataset\cite{meyerAutomotiveRadarDataset2019}. Additionally, SMURF\cite{liuSMURFSpatialMultiRepresentation2023} incorporates a point-wise kernel density estimation (KDE) branch, which calculates the density of point clouds within several predefined distance ranges, offering a detailed understanding of point distribution. The derived density information is then concatenated with pillarized features, resulting in enhanced BEV features. By refining the initial data, these methods lay a strong foundation for more accurate and reliable downstream processing.


In the domain of model backbone architecture, the incorporation of spatial attention mechanisms has been acknowledged as an effective strategy to address the sparsity and irregular distribution of point clouds. Xu et al.\cite{xuRPFANet4DRaDAR2021} implemented a self-attention mechanism to extract global information from the pillarized radar point cloud. Further advancing this concept, Shi et al.\cite{shiSMIFormerLearningSpatial2023} augment different view features using a combination of self-attention and cross-attention mechanisms. While self-attention focuses on understanding the relationships within a single view, cross-attention extends this understanding across different views, thus enabling a more comprehensive and integrated feature representation. Yan et al.\cite{yanMVFANMultiViewFeature2023} take a different approach by utilizing the attention matrix inherent in the self-attention mechanism to differentiate and reweight foreground and background points and their respective features. They also introduce a binary classification auxiliary loss to aid the learning process. Additionally, several studies have used spatial attention to fuse multimodal sensor data, not only addressing the noise and sparsity in radar point clouds, but also leveraging the strengths of different sensor modalities. These methods will be further elaborated in Section \ref{Sec:Percept-fusion}.




\subsection{Pre-CFAR Feature Extraction}\label{Sec:Percept-preCFAR}


In millimeter-wave (mmWave) radar signal processing, side lobe suppression and CFAR algorithms play a crucial role in reducing noise and minimizing false alarms. These techniques help extract signal peaks, thereby reducing data volume and computational cost. However, a consequential drawback of this approach is the sparsity of radar point clouds, characterized by diminished resolution. Given the profound advancements in deep learning, particularly in the processing of dense image data, a pivot in research focuses towards Pre-CFAR data is observed, aiming to utilize more underlying hidden information.


To our best knowledge, there are currently several datasets containing 4D mmWave radar Pre-CFAR data \cite{paekKRadar4DRadar2022, rebutRawHighDefinitionRadar2022, madaniRadatronAccurateDetection2022, nowruziDeepOpenSpace2020, zhangRADDetRangeAzimuthDopplerBased2021}. Notably, a subset of these datasets \cite{madaniRadatronAccurateDetection2022, nowruziDeepOpenSpace2020, zhangRADDetRangeAzimuthDopplerBased2021} have relatively lower elevation resolution, exceeding 15 degrees. Consequently, our survey will focus on the methodologies employed within the high resolution datasets provided by \cite{paekKRadar4DRadar2022, rebutRawHighDefinitionRadar2022}. This section intends to explain the comprehensive pipeline and the enhancements tailored for optimizing 4D mmWave radar characteristics.


As discussed in Section \ref{subsec:workflow}, the 4D mmWave radar signal processing workflow applies  FFT methodologies on raw ADC data to discretely process four dimensions: range, Doppler, azimuth, and elevation. This processing generate diverse data representations, including the Range-Doppler (RD) map, Range-Azimuth (RA) map, Range-Azimuth-Doppler (RAD) cube, and ultimately, a 4D tensor. From 2D maps to 4D tensors, the complexity of extracted features varies. Higher dimensional data requires more memory and computation for feature extraction. The extracted features are typically aligned to the RA axis in BEV under polar coordinates or the XY axis in Cartesian coordinates, connecting to detection or segmentation heads, serving as the foundational elements for subsequent detection or segmentation operations. 

\subsubsection{4D Tensor}

\begin{figure}[htbp]
\centerline{\includegraphics[width = \linewidth]{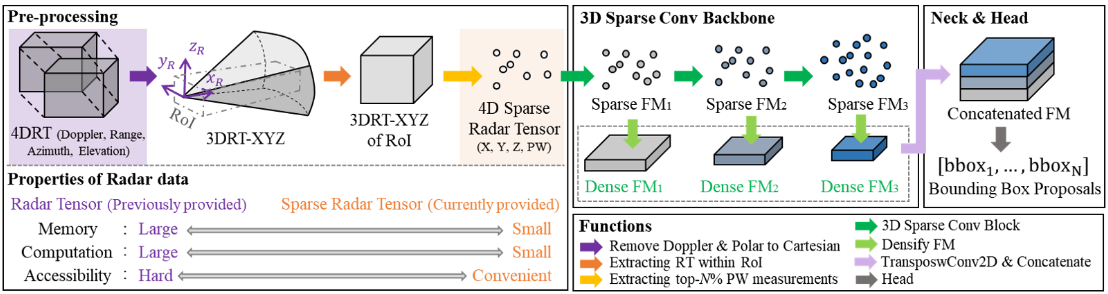}}
\caption{Overview of RTNH\cite{paekEnhancedKRadarOptimal2023} with 4D mmWave radar tensors}
\label{Fig:ERTNH}
\end{figure}


For the 4D tensor, Paek et al. \cite{paekEnhancedKRadarOptimal2023, paekKRadar4DRadar2022} opt to further extract a sparse tensor aligned to the Cartesian coordinate system, subsequently utilizing 3D sparse convolution to extract multi-scale spatial features. Experiments have demonstrated that retaining only the top-5\% elements with the highest power measurements can maintain detection accuracy while significantly enhancing processing speed. Furthermore, the elevation information included in 4D tensors is essential facing 3D but not BEV 2D object detection.

\subsubsection{RAD Cube}

In the context of RAD cube processing, TransRadar \cite{dalbahTransRadarAdaptiveDirectionalTransformer2024} projects the data onto the AD, RD, and RA planes, respectively, and innovatively designs an adaptive directional attention block to encode features separately. 

\subsubsection{RD Map}

Works related to the RD map \cite{rebutRawHighDefinitionRadar2022, girouxTFFTRadNetObjectDetection2023, bootEfficientDASHAutomatedRadar2023} generally encode features along the RD dimensions using CNN or Swin Transformer structures. Subsequently, a noteworthy technique involves the transposition of the Doppler dimension with the channel dimension, thereby redefining the conventional channel axis as the azimuth axis, followed by a series of deconvolution and upsampling steps to extrapolate features defined along the range-azimuth axes. 

\subsubsection{RA Map}

RA map data, inherently aligned with the polar coordinate system, is amenable to direct processing through dense feature extraction networks. The generated BEV features are then either converted into the Cartesian coordinate system via bi-linear interpolation \cite{madaniRadatronAccurateDetection2022} or utilized within polar-based detection frameworks \cite{liuEchoesPointsUnleashing2023}. 


\begin{figure}[htbp]
\centerline{\includegraphics[width = \linewidth]{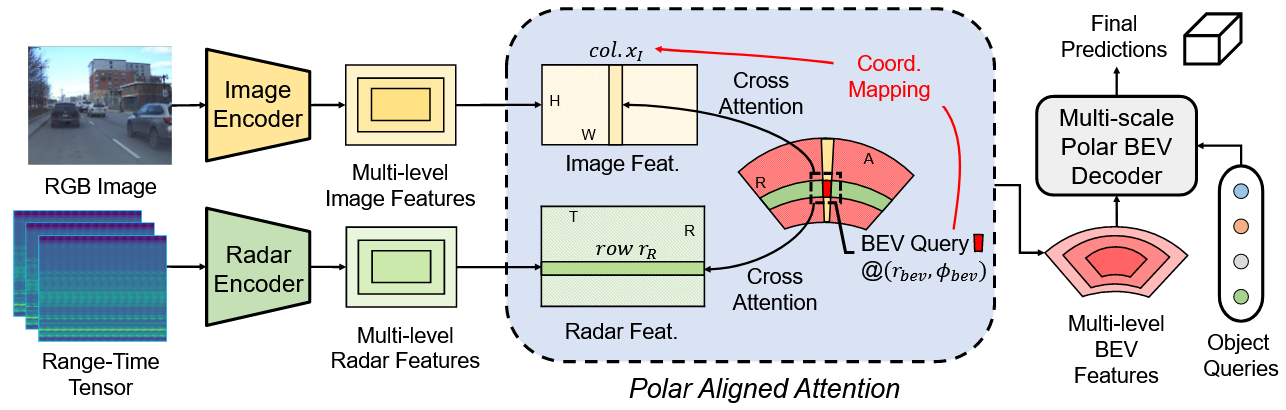}}
\caption{Overview of Echofusion \cite{liuEchoesPointsUnleashing2023} with RT maps and images}
\label{Fig:echofusion}
\end{figure}

\subsubsection{Raw ADC Data}

Recently, some studies have also shifted towards addressing the elevated computational demands by performing FFT on raw ADC data. Consequently, strategies have emerged wherein raw ADC data is directly processed via complex-valued linear layers\cite{girouxTFFTRadNetObjectDetection2023}, utilizing the prior knowledge of the Fourier transform. Alternatively, Liu et al.\cite{liuEchoesPointsUnleashing2023} have leveraged data derived from a single-range FFT operation to generate Range-Time (RT) representations. Comparative analyses show the performance gap between the RT map and RA map with camera modality is within the error bar. These findings suggest that the resolution of azimuth angles and the permutation of Doppler-angle dimensions, as previously posited by Rebut et al.\cite{rebutRawHighDefinitionRadar2022} and Giroux et al.\cite{girouxTFFTRadNetObjectDetection2023}, may not be requisite for achieving satisfactory performance outcomes.



\subsection{Multi-Modal Fusion Methods}\label{Sec:Percept-fusion}


Considering the capability of 4D mmWave radars to furnish point cloud data, several scholars have embarked on integrating this information with inputs from cameras or LiDAR systems to enhance the accuracy and robustness of the perception model. Generally, there are three fusion levels for different modalities: data level, feature level, and decision level. Existing research primarily focuses on the feature-level fusion.

\subsubsection{4D Radar with Vision}

As for 4DRV (4D mmWave Radar and Vision) fusion, 4D mmWave radars offer the ability to deliver high-precision depth and velocity information in a cost-effective manner, thereby mitigating the limitations inherent in camera systems and enhancing the accuracy of 3D detection. In recent studies, 4D mmWave radar signals are typically transformed into 2D image-like features, facilitating their practical deployment in conjunction with camera images. This fusion strategy leverages the strengths of both modalities, enabling a more comprehensive and accurate representation of the environment for advanced perception tasks.

Exemplifying this approach, Meyer et al.\cite{meyerDeepLearningBased2019} adapt a CNN architecture initially designed for camera-LiDAR fusion \cite{kuJoint3DProposal2018} to process RGB images with height and density maps generated from 4D mmWave radar point clouds. Remarkably, their fusion network demonstrates enhanced precision employing  radar instead of LiDAR point clouds, achieving an average precision (AP) of 61\% on the Astyx dataset \cite{meyerAutomotiveRadarDataset2019}. A subsequent study is performed by Cui et al.\cite{cui3DDetectionTracking2021} with a novel self-supervised model adaptation block\cite{valadaSelfSupervisedModelAdaptation2020}, which dynamically adapts the fusion of different modalities in accordance with the object properties. Besides, a FV map is generated from the 4D mmWave radar point clouds together with the BEV image. The presented method outperforms the former study\cite{meyerDeepLearningBased2019} by up to 9.5\% in 3D AP. The FV map effectively leverages the elevation information provided by 4D mmWave radars and achieves easier fusion with the monocular camera feature, balancing detection accuracy and computational efficiency.


Additionally, recent works such as RCFusion\cite{zhengRCFusionFusing4D2023} and LXL\cite{xiongLXLLiDARExcluded2023} have advanced the integration of attention mechanisms for the fusion of image and 4D mmWave radar features. They begin by separately extracting BEV features from the camera and radar branches, then employ convolutional networks to create scale-consistent attention maps, effectively delineating the occupancy grid for target objects. The distinction lies in the fact that RCFusion\cite{zhengRCFusionFusing4D2023} generates 2D attention maps in both the camera and radar branches, which are then multiplied with the BEV feature from the other modality. Conversely, LXL\cite{xiongLXLLiDARExcluded2023} solely utilizes the radar BEV feature to infer the 3D occupancy grid, which is then multiplied with the 3D image voxel features to achieve attention sampling of the image features.

\subsubsection{4D Radar with LiDAR}

Despite the notable advantages of 4DRV fusion, the vision-based branch may still struggle when facing aggressive lighting changes or adverse weather conditions, which in turn affects the overall performance of the model. Addressing this challenge, Wang et al. \cite{wangInterFusionInteractionbased4D2022} first explore the advantages of 4DRL(4D mmWave Radar and LiDAR) fusion with an interaction-based fusion framework. They design an InterRAL (Interaction of Radar and LiDAR) module and update pillars from both modalities to enhance feature expression. The efficacy of this approach is substantiated through a series of ablation experiments, demonstrating the potential of this fusion strategy in improving the robustness and performance of perception models under challenging conditions.

\begin{figure}[htbp]
\centerline{\includegraphics[width = \linewidth]{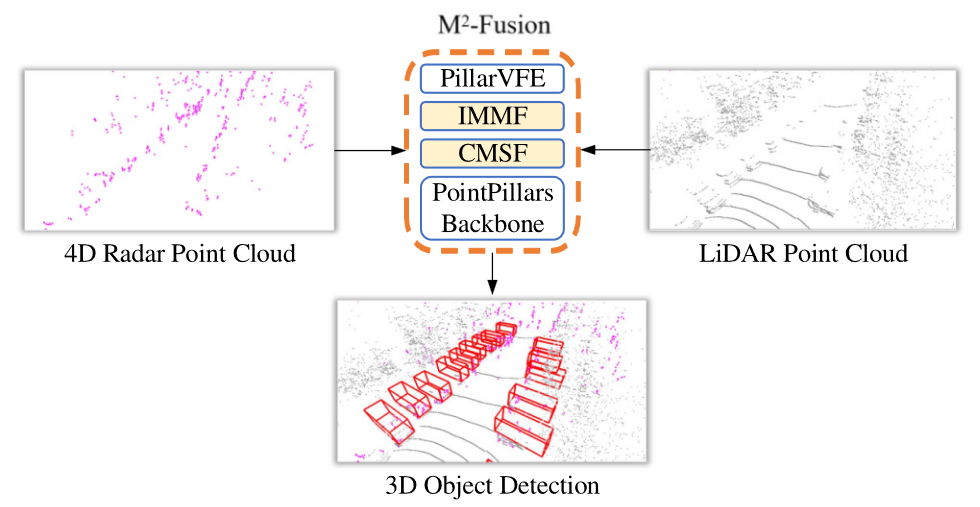}}
\caption{Overview of the $M^2$ 4DRL fusion model \cite{wangMultiModalMultiScaleFusion2023}}
\label{Fig:RLFusion}
\end{figure}

In a subsequent investigation, Wang et al. \cite{wangMultiModalMultiScaleFusion2023} propose the $M^2$-Fusion network that integrates an interaction-based multi-modal fusion(IMMF) block and a center-based multi-scale fusion(CMSF) block. Evaluated in the Astyx dataset\cite{meyerAutomotiveRadarDataset2019}, this novel approach outperforms mainstream LiDAR-based object detection methods significantly. As LiDARs can accurately detect objects at close range, 4D mmWave radars have a greater detection range owing to its penetrability, the fusion of 4DRL presents a promising technical solution that combines cost-effectiveness with high-quality performance.

\subsection{Challenge}\label{Sec:Percept-challenge}


Current methodologies for 4D mmWave radar point cloud perception predominantly adapt established techniques from LiDAR applications, whereas approaches for pre-CFAR data often draw from the vision domain. Although the data formats exhibit similarities, the unique characteristics of mmWave radar data, specifically Doppler velocity and intensity information, warrant more focused attention for effective feature extraction. Moreover, pre-CFAR radar data characteristically contains a significantly higher ratio of background to actual objects (foreground)\cite{dalbahTransRadarAdaptiveDirectionalTransformer2024}, a factor that complicates data interpretation and model training. The inherent noise within radar data further presents a substantial challenge for learning algorithms. The resilience of 4D mmWave radar models under out-of-distribution (OoD) conditions also remains inadequately explored and understood\cite{paekKRadar4DRadar2022}, which underscores the necessity for refined methodologies that can more accurately account for the distinct properties of mmWave radar data, thereby enhancing model robustness and performance in real-world scenarios.

\section{SLAM Applications} \label{Sec:SLAM}
In challenging environments where satellite-based positioning is unreliable or high-definition maps are absent, localization and mapping by perception sensors becomes indispensable. Recently, a collection of SLAM research has been conducted utilizing the emerging 4D mmWave radars.

\begin{figure*}
    \centering
    \includegraphics[width = \linewidth]{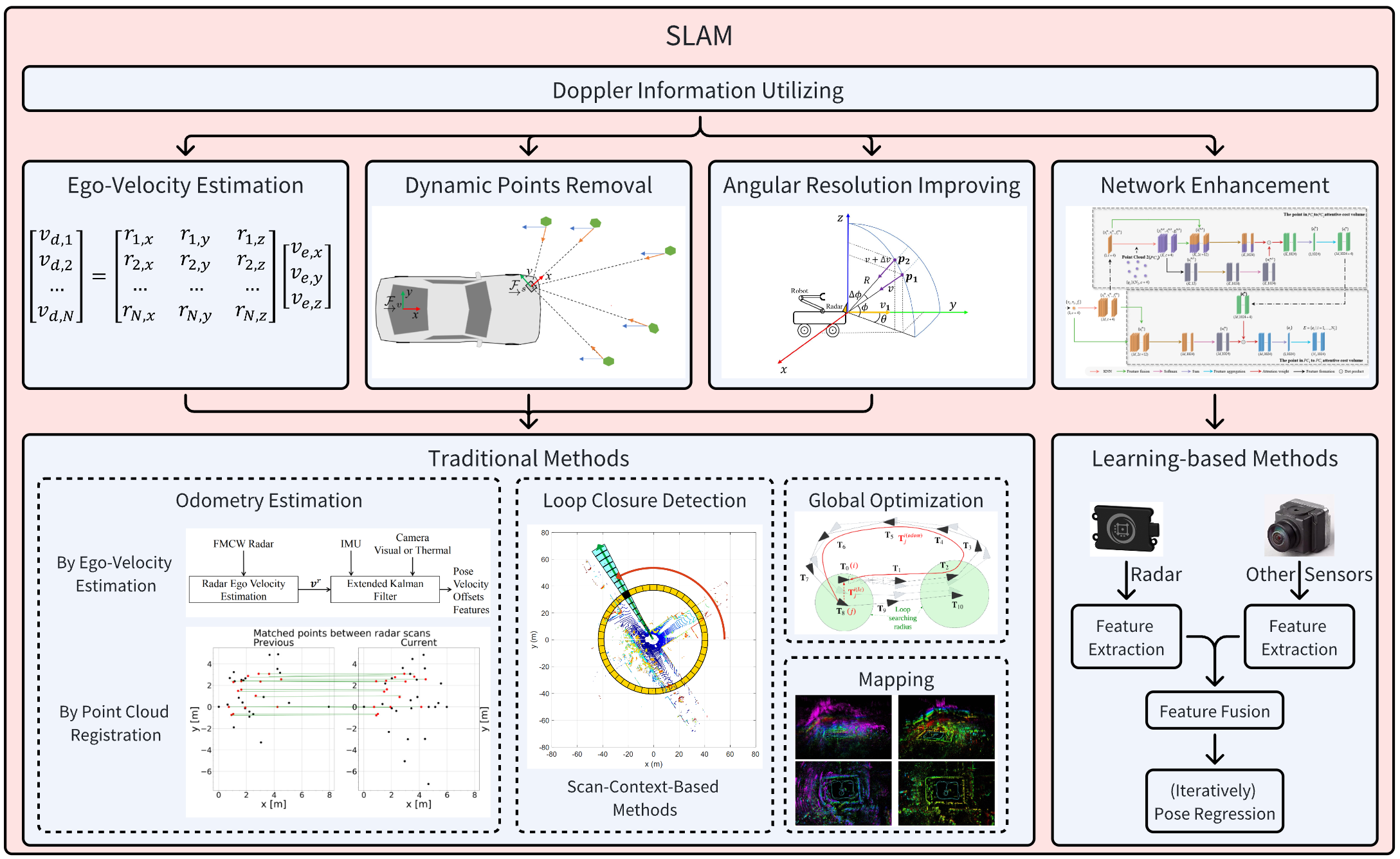}
    \caption{The taxonomy of 4D mmWave radar-based SLAM. Example images are from \cite{zhuang4DIRIOM4D2023, ngContinuoustimeRadarinertialOdometry2022, chengNovelRadarPoint2022, luEfficientDeepLearning4D2023, doerRadarVisualInertial2021, michalczykTightlyCoupledEKFBasedRadarInertial2022, kimScanContextEgocentric2018, zhang4DRadarSLAM4DImaging2023, luMilliEgoSinglechipMmWave2020}}
    \label{Fig:SLAM}
\end{figure*}

As Fig. \ref{Fig:SLAM} demonstrates, the unique Doppler information contained within radar point clouds presents a notable opportunity for exploitation, which will be discussed in Section \ref{Sec:SLAM-Doppler}. Subsequently, both traditional and learning-based SLAM approaches will be introduced in Section \ref{Sec:SLAM-Traditional} and Section \ref{Sec:SLAM-Learning}, respectively. And Section \ref{Sec:SLAM-Challenge} briefly introduces current challenges in 4D mmWave radar-based SLAM.

\subsection{Doppler Information Utilizing} \label{Sec:SLAM-Doppler}

The Doppler information constitutes a significant advantage of 4D mmWave radars, especially in the context of SLAM applications. Generally speaking, the utilization of Doppler information can be categorized into several categories:

\subsubsection{Ego-Velocity Estimation}
To estimate the ego-velocity of radar using Doppler information, a straightforward approach is the linear least squares (LSQ). Doppler velocity reflects the radial relative velocity between the object and the ego vehicle. Consequently, it is only the Doppler velocity from stationary objects that  are viable for deducing the ego vehicle's velocity, while Doppler information from dynamic objects is considered as outliers. Under the assumption that the majority of surroundings points are from stationary objects, LSQ offers a suitable and computationally efficient method for ego-velocity estimation \cite{doerEKFBasedApproach2020}. 

Assuming  the 3D spatial coordinates of a point in the 4D mmWave radar system are denoted as $\boldsymbol{p}_i$. Its directional vector is determined as follows:

\begin{equation}
    \boldsymbol{r}_i=\frac{\boldsymbol{p}_i}{\|\boldsymbol{p}_i\|}.
\end{equation}

Let us consider a scenario where point $\boldsymbol{p}_i$ is detected from a stationary object, the ideally measured Doppler velocity $v_{d,i}$ represents the projection of the ego-velocity $\boldsymbol{v}_e$ onto the  line-of-sight vector connecting the 4D mmWave radar with point $\boldsymbol{p}_i$. This velocity can be mathematically expressed as follows:

\begin{equation}
    v_{d,i}=\boldsymbol{v}_e\cdot\boldsymbol{r}_i=v_{e,x}r_{i,x}+v_{e,y}r_{i,y}+v_{e,z}r_{i,z}.
\end{equation}

For a set of $N$ points constituting a single frame of the point cloud, the above equation can be generalized into a matrix formulation:

\begin{equation}
    \begin{bmatrix}
    v_{d,1}\\
    v_{d,2}\\
    ...\\
    v_{d,N}
    \end{bmatrix}=\begin{bmatrix}
    r_{1,x}&r_{1,y}&r_{1,z}\\
    r_{2,x}&r_{2,y}&r_{2,z}\\
    ...&...&...\\
    r_{N,x}&r_{N,y}&r_{N,z}
    \end{bmatrix}
    \begin{bmatrix}
    v_{e,x}\\
    v_{e,y}\\
    v_{e,z}
    \end{bmatrix}.
\end{equation}

Leveraging LSQ, an optimal solution can be obtained as follows:

\begin{equation}
    \boldsymbol{v}_e = \begin{bmatrix}
    v_{e,x}\\
    v_{e,y}\\
    v_{e,z}
    \end{bmatrix}=(\boldsymbol{R}^T\boldsymbol{R})^{-1}\boldsymbol{R}^T\begin{bmatrix}
    v_{d,1}\\
    v_{d,2}\\
    ...\\
    v_{d,N}
    \end{bmatrix},
\end{equation}

where

\begin{equation}
    \boldsymbol{R}=\begin{bmatrix}
    r_{1,x}&r_{1,y}&r_{1,z}\\
    r_{2,x}&r_{2,y}&r_{2,z}\\
    ...&...&...\\
    r_{N,x}&r_{N,y}&r_{N,z}
    \end{bmatrix}.
\end{equation}

Indeed, relying solely on the LSQ method may result in substantial inaccuracies, primarily due to the influence of dynamic objects and noise. To mitigate these effects, specific research endeavors have implemented techniques such as RANSAC to effectively remove dynamic points before the application of LSQ \cite{li4DRadarBasedPose2023}. 

An innovative enhancement to the conventional LSQ involves the incorporation of weighting mechanisms. Galeote-Luque et al. \cite{galeote-luqueDoppleronlySinglescan3D2023} weight each point by its signal power to diminish the influence of noise. Addressing the challenge posed by dynamic objects, Zhuang et al. \cite{zhuang4DIRIOM4D2023} propose a reweighted least squares method for the estimation of ego-velocity. The objective function is constructed as follows:

\begin{equation}
    \min\limits_{\boldsymbol{v}_e} \sum_{i=1}^{n} \lambda_i || v_{d,i}-\boldsymbol{r}_i \cdot \boldsymbol{v}_e||,
\end{equation}

where $\lambda_i$ denotes the weight of the $i$-th radar point. In the first iteration, $\lambda_i=1$, and $\lambda_i=1/(|| v_{d,i}-\boldsymbol{r}_i \cdot \boldsymbol{v}_e||+\epsilon), \epsilon=0.00001$. $\lambda_i$ in the subsequent iterations, which quantifies the difference between the actual Doppler velocity $v_{d,i}$ and the ideal Doppler velocity assuming the $i$-th point originates from a stationary object,denoted as $\boldsymbol{r}_i \cdot \boldsymbol{v}_e$. Through iterative refinement, the weights of points from dynamic objects are progressively decrease, culminating in a more accurate computation of the ego-velocity $\boldsymbol{v}_e$. The RCS values are then employed to weight point cloud registration residuals to reduce the impact of matches with large RCS differences \cite{wang4DRADARIMU2023}.

\subsubsection{Dynamic Points Removal}
Beyond ego-velocity estimation, another coherent exploration of Doppler information is to remove dynamic points, particularly leveraging the result derived from ego-velocity estimation. Zhang et al. \cite{zhang4DRadarSLAM4DImaging2023} \cite{zhang4DRTSLAMRobustSLAM2023} apply a RANSAC-like method, while Zhuang et al. \cite{zhuang4DIRIOM4D2023} utilize the weights to distinguish dynamic points in accordance with ego-velocity estimation.

\subsubsection{Angular Resolution Improving}
The angular resolution of 4D mmWave radars is determined by virtual TX-RX pairs mentioned in Section \ref{Sec:Theory}, yet the range and Doppler resolution are dictated by the frequency disparity between transmitted and received signals. Therefore, 4D mmWave radars typically exhibit superior performance in terms of range and Doppler resolution, as opposed to angular resolution. Cheng et al. \cite{chengNovelRadarPoint2022} demonstrate that for two points from stationary objects, provided they share the same range and azimuth, the differences in elevation $\Delta \phi$ and Doppler $\Delta v$ between the two points are interconnected. Similarly, the Doppler velocity resolution can also be converted into azimuth resolution. Hence, leveraging Doppler information can improve the angular resolution in particular azimuth and elevation ranges.

Drawing a parallel, Chen et al. \cite{chenDRIORobustRadarInertial2023} harness Doppler information to refine the point cloud and implement radar-inertial odometry using ground points, which exhibit stability in dynamic environments. Given that the resolution of the $z$ coordinate in point clouds is generally inferior than Doppler resolution, once a point is identified as a ground point, its $z$ coordinate can be recalculated using Doppler information and $x, y$ coordinates. Adopting a strategy inspired by RANSAC, the authors hypothesize and refine ground points, then estimate ego-velocity iteratively. The refined point clouds can subsequently be applied in other tasks such as object detection.

\subsubsection{Network Enhancement}
In the context of learning-based SLAM, Doppler information also holds significant value.As previously discussed, Doppler velocity can serve as an indicator of whether a point originates from a stationary or dynamic object, which enlightens authors of \cite{luEfficientDeepLearning4D2023} and \cite{zhuo4DRVONetDeep4D2023} to establish a velocity-aware attention module. This module leverages Doppler information to learn attention weights, thereby distinguishing between stationary and dynamic points.

\subsection{Traditional Methods} \label{Sec:SLAM-Traditional}
Traditional SLAM refers to methods with no neural networks, and can be composed into four modules: odometry estimation, loop closure detection, global optimization and mapping. Considering the unique characteristic of 4D mmWave radars, related research mostly feature in the former two modules, and will be discussed below.

\subsubsection{Odometry Estimation}
Odometry estimation is the core of localization and serves as a crucial component of SLAM. A substantial body of traditional research on odometry estimation has been conducted in the context of 4D mmWave radars.

Considering the inherent noise and sparsity of 4D mmWave radar point clouds, original odometry estimation research have primarily concentrates on the estimation of ego-velocity derived from the Doppler information instead of point cloud registration. Doer and Trommer make plenty of contributions to this field using Unmanned Aerial Vehicles (UAV). They fuse ego-velocity estimated by LSQ with the IMU data to perform UAV odometry estimation\cite{doerEKFBasedApproach2020, doerYawAidedRadar2021}, and further extend their work to multiple radars \cite{doerXRIORadarInertial2021, doerGNSSAidedRadar2022} and radar-camera fusion systems \cite{doerRadarVisualInertial2021}. As early investigations, these research efforts exhibit certain limitations. They rely on Manhattan world assumptions and consider the surroundings to be stationary, which may restrict the applicability in challenging outdoor scenarios.

Additionally, the ego-velocity estimated by 4D mmWave radar Doppler information has been explored by various researchers, in conjunction with additional assumptions, to achieve radar-based odometry. Ng et al. \cite{ngContinuoustimeRadarinertialOdometry2022} present a continuous-time framework that fuse the ego-velocity from multiple radars with the measurement of an IMU. The continuity of this framework facilitates closed-form expressions for optimization and makes it well-suited for asynchronous sensor fusion. Given the relatively low elevation resolution of 4D mmWave radars in contrast to the more precise Doppler information, Chen et al.\cite{chenDRIORobustRadarInertial2023} propose a method to detect ground points and estimate ego velocity iteratively. Furthermore, Galeote-Luque et al. \cite{galeote-luqueDoppleronlySinglescan3D2023} combine the linear ego-velocity estimated from radar point clouds with the kinematic model of the vehicle to reconstruct the 3D motion of the vehicle. 

\begin{figure}[htbp]
\centerline{\includegraphics[width = \linewidth]{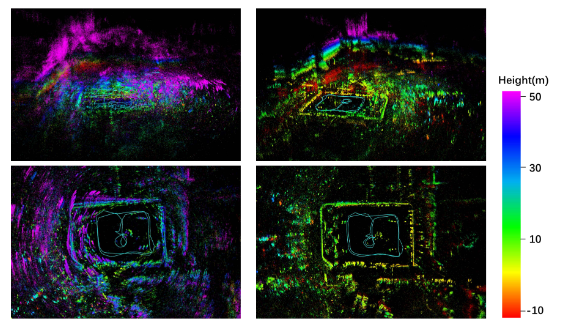}}
\caption{The mapping performance of 4D iRIOM \cite{zhuang4DIRIOM4D2023}}
\label{Fig:4DIROM}
\end{figure}

Recent research has shifted focus from direct odometry estimation through Doppler-based ego-velocity, as seen in above studies, to point cloud registration akin to traditional LiDAR odometry. In 4D mmWave radar point cloud registration, Doppler information utilization and noise and sparsity handling are two main concerns. Michalczyk et al. \cite{michalczykTightlyCoupledEKFBasedRadarInertial2022} pioneer the realization of 3D point registration across sparse and noisy radar scans based on the classic Hungary algorithm \cite{kuhnHungarianMethodAssignment1955}. Additionally, emerging research in 4D mmWave radar SLAM is also delving into specialized designs of point cloud registration techniques. Zhuang et al. \cite{zhuang4DIRIOM4D2023} develop a 4D mmWave radar inertial odometry and mapping system named 4D iRIOM employing iterative Extended Kalman Filter (EKF). To mitigate the effects of sparsity, they introduce an innovative point cloud registration method between each scan and submap. This method accounts for the local geometry of points in the current scan and the corresponding $N$ nearest points in the submap, weighting the distances between them by their covariance to achieve a distribution-to-multi-distribution effect. The results of this approach are shown in Fig. \ref{Fig:4DIROM}. Moreover, they incorporate 4D iRIOM with GNSS and propose G-iRIOM \cite{wang4DRADARIMU2023}, which further utilizes RCS value to weight the point cloud registration. Besides, the pose graph optimization is applied by Zhang et al.\cite{zhang4DRadarSLAM4DImaging2023, zhang4DRTSLAMRobustSLAM2023} to construct a 4D mmWave radar SLAM system adapted from a well-known LiDAR SLAM method named hdl\_graph\_slam \cite{koidePortableThreedimensionalLIDARbased2019}. Building upon the traditional point cloud registration algorithm Generalized Iterative Closest Point (GICP)\cite{segalGeneralizedicp2009}, they propose an adaptive probability distribution-GICP, assigning different covariance to each point according to its uncertainty inferred from the coordinate of each point, given that points at greater distances may exhibit increased uncertainty. This design considers not only the geometric distribution of neighboring points but also the spatial variance of each point. Also considering point uncertainty, Li et al \cite{li4DRadarBasedPose2023} propose 4DRaSLAM to incorporate the probability density function of each point to develop a probability-aware Normal Distributions Transform (NDT) \cite{biberNormalDistributionsTransform2003} routine for scan-to-submap point cloud registration. Notably, the ego-velocity estimated from 4D mmWave radar Doppler information is utilized as a pre-integration factor in this system to replace the role of IMU.

\subsubsection{Loop Closure Detection}
With respect to loop closure detection, relevant inventive research remains scarce. Existing 4D mmWave radar SLAM research that involve loop closure detection \cite{zhuang4DIRIOM4D2023, li4DRadarBasedPose2023, zhang4DRadarSLAM4DImaging2023} typically reference the well-known Scan Context algorithm \cite{kimScanContextEgocentric2018}. The original Scan Context algorithm partitions a LiDAR point cloud into several bins based on the azimuth, utilizing the maximum height of the points in each bin to encode the entire point cloud into an image. However, considering the relatively low resolution of height information provided by 4D mmWave radars, maximum intensity instead of height is adapted as context for loop closure detection in these systems. 

\subsection{Learning-based Methods} \label{Sec:SLAM-Learning}

Research on learning-based 4D mmWave radar SLAM has predominantly focused on odometry estimation, replacing traditional point cloud registration and pose regression by deep networks.

As the originator, Lu et al. \cite{luMilliEgoSinglechipMmWave2020} design CNNs and Recurrent Neural Networks (RNNs) to extract the features from radar point clouds and IMU data, respectively. Subsequently, they propose a two-stage cross-modal attention mechanism to achieve feature integration. An RNN is utilized additionally to capture the long-term temporal dynamics of the system. 

\begin{figure}[htbp]
\centerline{\includegraphics[width = \linewidth]{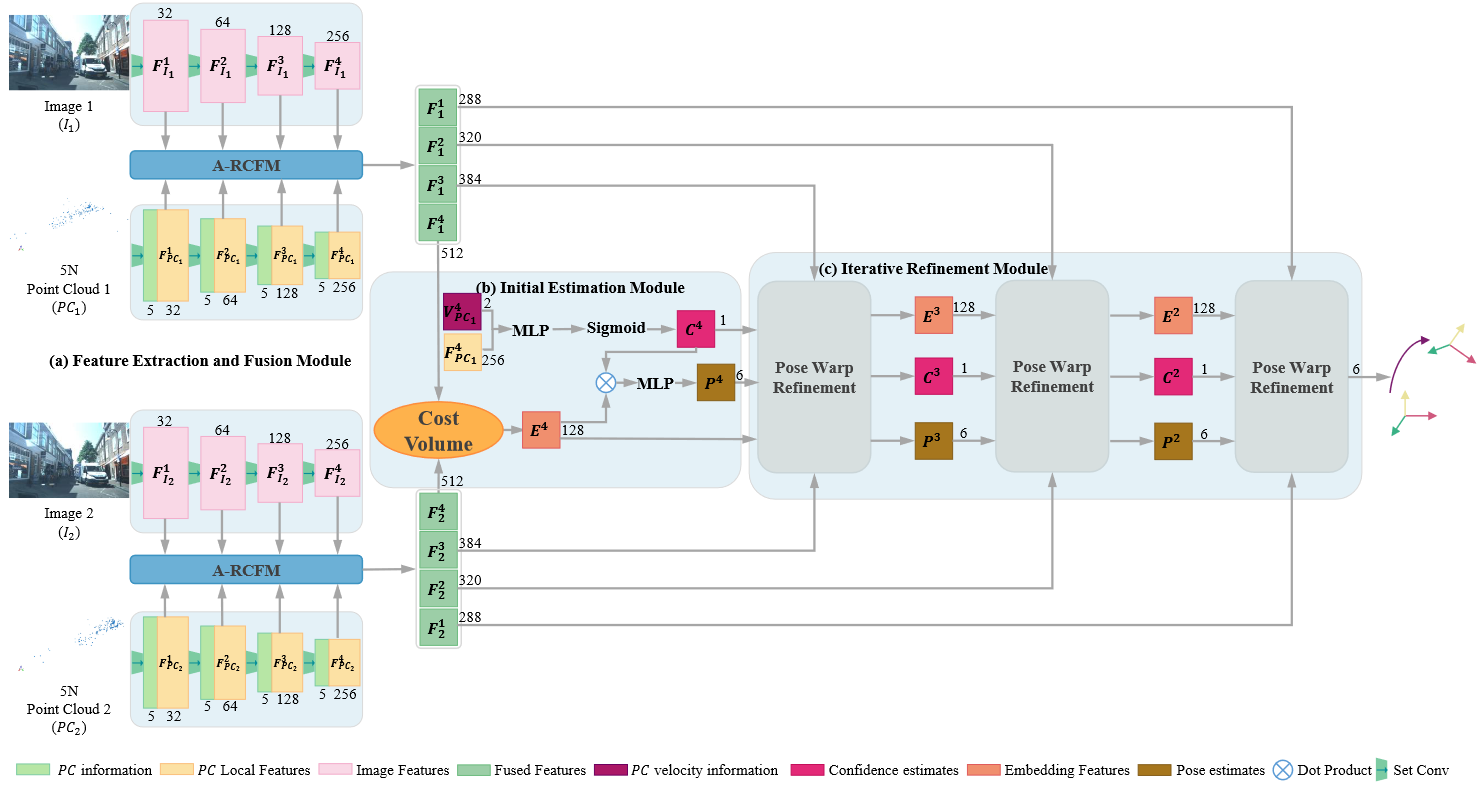}}
\caption{The data fusion and localization pipeline of 4DRVO-Net \cite{zhuo4DRVONetDeep4D2023}}
\label{Fig:4DRVO}
\end{figure}

To make full use of Doppler information, 4DRO-Net \cite{luEfficientDeepLearning4D2023} establishes a velocity-aware attention cost volume network within a coarse-to-fine hierarchical optimization framework to iteratively estimate and refine pose estimation. Global-level and point-level features are extracted to generate initial pose estimations and subsequent corrections. 4DRVO-Net \cite{zhuo4DRVONetDeep4D2023} further extracts and fuses image features with 4D mmWave radar point cloud features. Fig. \ref{Fig:4DRVO} displays the pipeline of 4DRVO-Net. The adaptive fusion module employs a deformable attention-based spatial cross-attention mechanism to align each 4D mmWave radar feature with corresponding image feature to achieve optimal fusion.

All these advanced learning-based radar odometry methods perform odometry estimation in an end-to-end fashion, i.e., the network ingests 4D mmWave radar point clouds (supplemented with images for fusion methods), and directly outputs odometry estimation. This benefits research on comprehensive end-to-end autonomous driving systems.

\subsection{Challenge} \label{Sec:SLAM-Challenge}

The Doppler velocity inherent in 4D mmWave radar point clouds has been recognized and utilized in SLAM to realize ego-velocity estimation and dynamic object removal, etc. However, considering adequate semantic information provided by 4D mmWave radars, such as intensity, traditional 3D registration methods like ICP may be less effective. Therefore, the exploration of learning-based methods or feature-based registration could yield more effective results. The optimal exploitation of these semantic information within the autonomous driving sphere remains a relatively open question.

Given that radar point clouds are significantly less data-heavy than their tensor counterparts, and methodologies developed for LiDARs can be adapted to 4D mmWave radar point clouds with minimal modifications, the majority of SLAM studies preferentially employ radar point clouds as their input data instead of radar tensors. But regarding mapping, the sparsity of 4D mmWave radar point clouds presents a significant challenge. A potential solution could lie in mapping and rendering the environment using 4D tensor-level data.

\section{Datasets} \label{Sec:Datasets}

Public datasets play an indispensable role in the advancement of 4D mmWave radar-based algorithms, as they furnish essential platforms for the development, benchmarking, and comparative analysis of diverse algorithms, thereby stimulating research in the field. This section categorizes and introduces current available datasets containing 4D mmWave radars, which are summarized in Table \ref{Tab:Datasets}.

\begin{table*}[!htbp]
    \centering
    \caption{4D mmWave radar datasets}
    \resizebox{\linewidth}{!}{
    \begin{tabular}{cccccccccccc} 
        \toprule
        \multirow{2}{*}{Dataset} &\multicolumn{4}{c}{Resolution} & \multirow{2}{*}{\makecell[c]{Total\\ Frames}} & \multirow{2}{*}{\makecell[c]{Labeled\\ Frames}} &\multirow{2}{*}{\makecell[c]{Data \\ Formats$^{1}$}} &\multirow{2}{*}{Modality$^{2}$} &\multirow{2}{*}{Bounding box} &\multirow{2}{*}{\makecell[c]{Tracking \\ ID}} &\multirow{2}{*}{Odometry}\\
        \cmidrule{2-5} 
            & Azi. & Ele. & Range(m) & Velo.(m/s) \\
        \midrule
        \multicolumn{12}{c}{Datasets for Perception}\\
        \midrule
        Astyx \cite{meyerAutomotiveRadarDataset2019} &N/M$^{3}$ &N/M &N/M &N/M &0.5K &0.5K &RPC &RCL &3D &\checkmark &$\times$ \\
        RADIal \cite{rebutRawHighDefinitionRadar2022} &0.1$^{\circ}$ &1$^{\circ}$ &0.2 &0.1 &25K &8.3K &ADC, RAD, RPC &RCL &2D &$\times$ &\checkmark \\
        VoD \cite{palffyMultiClassRoadUser2022} &1.5$^{\circ}$ &1.5$^{\circ}$ &0.2 &0.1 &8.7K &8.7K &RPC &RCI &3D &\checkmark &\checkmark \\
        TJ4DRadSet \cite{zhengTJ4DRadSet4DRadar2022} &1$^{\circ}$ &1$^{\circ}$ &0.86 &N/M &40K &7.8K &RPC &RCL &3D &\checkmark &\checkmark \\
        K-Radar \cite{paekKRadar4DRadar2022} &1$^{\circ}$ &1$^{\circ}$ &0.46 &0.06 &35K &35K &4DRT, RPC &RCLI &3D &\checkmark &\checkmark \\
        Dual Radar \cite{zhangDualRadarMultimodal2023} &1.2$^{\circ}$ &2$^{\circ}$ &0.22 &N/M &50K &10K &RPC &RCL &3D &\checkmark &$\times$ \\
        SCORP \cite{nowruziDeepOpenSpace2020} &15$^{\circ}$ &30$^{\circ}$ &12 &0.33 &3.9K &3.9K &ADC, RAD, RPC &RC &$\times$ &$\times$ &\checkmark \\ 
        Radatron \cite{madaniRadatronAccurateDetection2022} &1.2$^{\circ}$ &18$^{\circ}$ &0.05 &N/M &152K &16K &RA &RC &2D &$\times$ &$\times$ \\
        \midrule
        \multicolumn{12}{c}{Datasets for SLAM$^{4}$} \\
        \midrule
        Coloradar \cite{kramerColoRadarDirect3D2022} &1$^{\circ}$ &22.5$^{\circ}$ &0.12 &0.25 &108K &- &ADC, RAE, RPC &RLI &$\times$ &$\times$ &\checkmark \\
        MSC-RAD4R \cite{choiMSCRAD4RROSBasedAutomotive2023} &1$^{\circ}$ &0.5$^{\circ}$ &0.86 &0.27 &90K &- &RPC &RCLI &$\times$ &$\times$ &\checkmark \\
        NTU4DRadLM \cite{zhangNTU4DRadLM4DRadarcentric2023} &0.5$^{\circ}$ &0.1$^{\circ}$ &0.86 &N/M &61K &- &RPC &RCLIT &$\times$ &$\times$ &\checkmark \\
        \bottomrule
    \end{tabular}}
    \label{Tab:Datasets}
    \begin{tablenotes}
        \item{$^{1}$ADC: raw radar data after Analog-to-Digital Converter; RA: Range-Azimuth map; RD: Range-Doppler map; RAD: Range-Azimuth-Doppler cube; \\
        RAE: Range-Azimuth-Elevation cube; 4DRT: Range-Azimuth-Elevation-Doppler Tensor; RPC: Radar Point Cloud;}
        \item{$^{2}$R: Radar, C:Camera, L:LiDAR, I:IMU(Inertial Measurement Unit), T:Thermal Camera}
        \item{$^{3}$N/M: Not Mentioned.}
        \item{$^{4}$ Datasets designed only for SLAM contain no labels like bounding box and tracking ID.}
    \end{tablenotes}
\end{table*}

\subsection{Datasets for Perception}

Datasets for 4D mmWave radar perception typically include 3D (or 2D) bounding boxes for object detection tasks, and tracking ID for object tracking tasks. Astyx \cite{meyerAutomotiveRadarDataset2019} represents the first 4D mmWave radar dataset. It consists of 500 synchronized frames (radar, LiDAR, camera) encompassing approximately 3,000 annotated 3D object annotations. As a pioneer dataset in the realm of 4D mmWave radars, the volume of data in Astyx is relatively limited. 

In order to facilitate researchers in handling radar data in a more fundamental manner, the RADIal dataset \cite{rebutRawHighDefinitionRadar2022} records raw radar data after Analog-to-Digital Converter, which serves as the foundation for generating various conventional radar representations, such as radar tensors and point clouds. Given that the raw ADC data is not interpretable by human eyes, the annotations in the RADIal dataset are presented as 2D bounding boxes in the image plane.

To advance 4D mmWave radar-based multi-class 3D road user detection, the VoD dataset \cite{palffyMultiClassRoadUser2022} is collected comprising LiDAR, camera, and 4D mmWave radar data. It contains 8693 frames of data captured in complex urban traffic scenarios, and includes 123106 3D bounding box annotations of both stationary and dynamic objects and tracking IDs for each annotated object. 

In a similar vein, the Tj4DRadSet dataset \cite{zhengTJ4DRadSet4DRadar2022} comprises 44 consecutive sequences, totaling 7757 synchronized frames, well-labeled with 3D bounding boxes and trajectory IDs. Notably, the TJ4DRadSet dataset offers occlusion and truncation indicators in each object to distinguish between different levels of detection difficulty. Unlike the VoD dataset, Tj4DRadSet is characterized by its inclusion of a broader and more challenging array of driving scenario clips, such as urban roads, highways, and industrial parks.

\begin{figure}[htbp]
\centerline{\includegraphics[width = \linewidth]{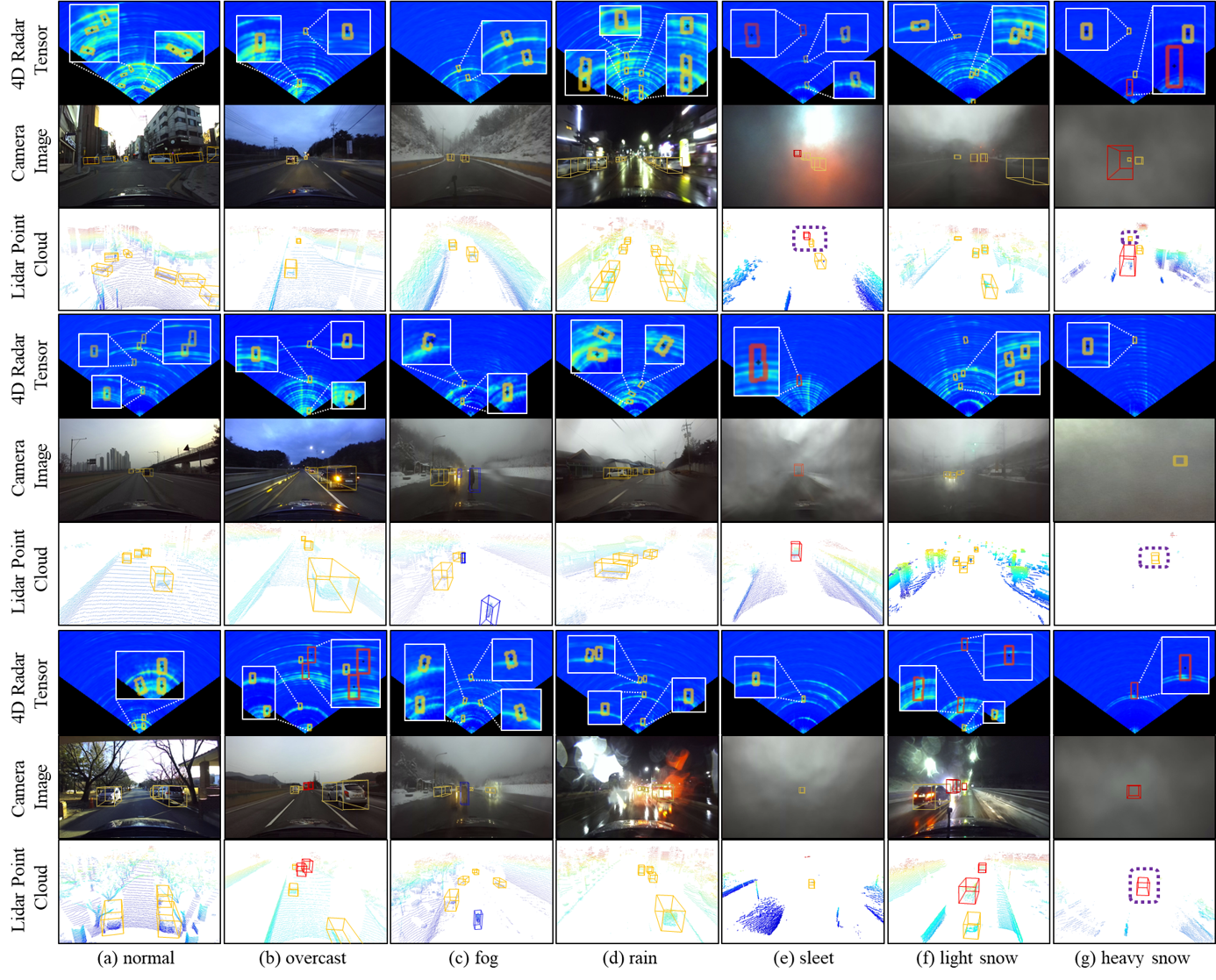}}
\caption{Data annotations in K-Radar dataset \cite{paekKRadar4DRadar2022} across different weather conditions.}
\label{Fig:KRadar}
\end{figure}

To the best of our knowledge, K-Radar dataset\cite{paekKRadar4DRadar2022} currently contains most diverse scenarios in 4D mmWave radar datasets as it collects 35k frame under a variety of weather conditions, including sunny, foggy, rainy, and snowy. K-Radar not only provides 4D mmWave radar data but also includes high-resolution LiDAR point clouds, surround RGB imagery from four stereo cameras, RTK-GPS and IMU data from the ego-vehicle. Fig. \ref{Fig:KRadar} illustrates the comparison of different modality sensors across different weather conditions. It is worth mentioning that K-radar is currently the only dataset that provides range-azimuth-elevation-Doppler tensors. 

\begin{figure}[htbp]
\centerline{\includegraphics[width = \linewidth]{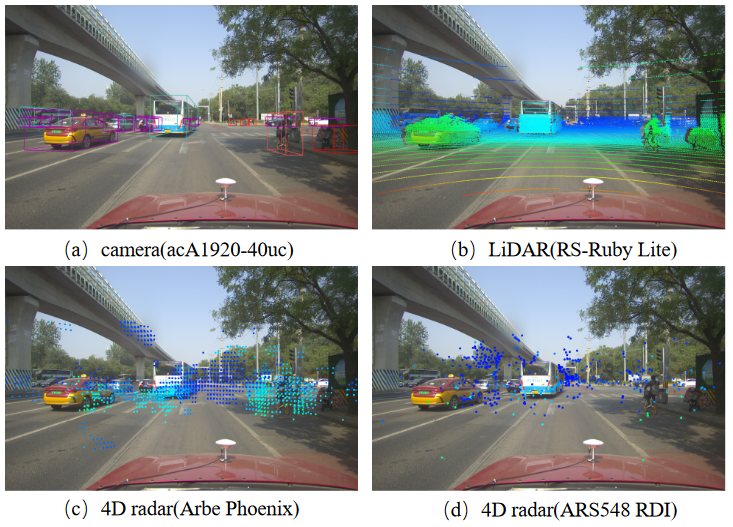}}
\caption{The data visualization of Dual Radar dataset. \cite{zhangDualRadarMultimodal2023}}
\label{Fig:DualRadar}
\end{figure}

However, each of the datasets mentioned above contains only one type of radar, making it challenging for researchers to analyze and compare the performance of different 4D mmWave radars. The recently unveiled Dual Radar dataset\cite{zhangDualRadarMultimodal2023}, as illustrated in Figure \ref{Fig:DualRadar}, encompasses two distinct types of mmWave radars, the Arbe Phoenix and the ARS548 RDI radar. Dual Radar enables an investigation into the impact of different sparsity levels in radar point clouds on object detection performance, providing assistance in the selection of radar products.

In addition, several datasets enumerated in Table \ref{Tab:Datasets} provide radar data with relatively low angular resolution \cite{nowruziDeepOpenSpace2020, madaniRadatronAccurateDetection2022}. These datasets, which are not detailed within the scope of this paper, provide alternative radar data characteristics that may be leveraged in different research contexts.

\subsection{Datasets for SLAM}

As its ability of perception in severe environments, the 4D mmWave radar enhances robust localization in difficult conditions, hence leading to the release of several 4D mmWave radar datasets specifically designed for localization, mapping and SLAM. Besides, any above dataset containing odometry information, not just those designed for SLAM, can also be employed for SLAM.

ColoRadar \cite{kramerColoRadarDirect3D2022} comprising approximately 2 hours of data from radar, LiDAR, and the 6-DOF pose ground truth. It provides radar data of three processing levels: raw ADC data, 3D range-azimuth-elevation tensors derived by compressing the Doppler dimension of 4D radar tensors, and radar point clouds. This dataset collects data from a variety of unique indoor and outdoor environments, thus providing a diverse spectrum of sensor data.

Considering SLAM in severe environments, the MSC-RAD4R dataset \cite{choiMSCRAD4RROSBasedAutomotive2023} records data under a wide range of environmental conditions, with the same route yielding data from both clear and snowy weather for comparison. Additionally, MSC-RAD4R introduces artificially generated smoke environment data generated by a smoke machine, further emphasizing the robust capabilities of 4D mmWave radars.

The NTU4DRadLM dataset \cite{zhangNTU4DRadLM4DRadarcentric2023} is a recent contribution to this field captured using both robotic and vehicular platforms. Distinguished from its predecessors, NTU4DRadLM delivers an extensive array of localization-related sensor data, including the 4D mmWave radar, LiDAR, camera, IMU, GPS, and even a thermal camera. Furthermore, it encompasses a wide range of road conditions, encompassing structured, semi-structured, and unstructured roads, spanning both small-scale environments (e.g., gardens) and large-scale urban settings.

\subsection{Challenge}

Taking into account the datasets mentioned above, it becomes apparent that there is a lack of labeled radar ADC data required for deep radar detection. To address this deficiency, using synthetic data generation techniques based on 4D mmWave radar sensor models can be considered, though radar modeling is challenging due to multiple effects of radar signal processing like multi-path reflections and signal interference.

Moreover, the scales of current 4D mmWave radar datasets far below other famous autonomous driving datasets such as nuScenes\cite{caesarNuScenesMultimodalDataset2020} and ONCE\cite{maoOneMillionScenes2021}. For the evaluation of algorithmic generalizability and for facilitating comparative analysis with other sensor modalities, large scale 4D mmWave datasets are indispensable in the future.

\section{Future Trends} \label{Sec:Future}
4D mmWave radars have the potential to bring about transformative advancements in the field of autonomous vehicles. Nonetheless, it is far from mature at the moment. The prospective evolution of 4D mmWave radar technology in autonomous driving is likely to be contingent upon advancements in several key domains.

\subsection{Noise and Sparsity Handling}


Despite the superior resolution of 4D mmWave radars compared to traditional 3D radars, factors such as antenna design, power, and multi-path effects still lead to significant noise and sparsity issues, impacting the safety of autonomous driving applications.

\subsubsection{Radar Data Generation}


In recent years, the computer vision field has seen numerous studies on image super-resolution, and the related theories and models can be transferred to the mmWave radar domain for application. Learning-based methods show promise for generating higher-resolution radar data, such as replacing traditional signal processing steps like CFAR and DBF, is a promising direction. However, considering the large spectral data volume of mmWave radar (a single frame 4D radar tensor in the K-RADAR dataset \cite{paekKRadar4DRadar2022} is around 200MB), related researches still necessitates improvements in real-time processing and addressing high bandwidth requirements for pre-CFAR data transmission. Compared to Transformer-based methods with quadratic time complexity, we believe that models using space state models \cite{guMambaLinearTimeSequence2023, liuVMambaVisualState2024} with linear complexity will have better prospects in this field.

\subsubsection{Application Algorithms Redesign}


Current mmWave radar perception algorithms often construct dense BEV features using pillars or cylinders, which may not be optimal for noisy, lower-resolution mmWave radar data. We believe a more efficient form is to organize object features in a sparse query manner and aggregate sensor data features using attention mechanisms, a paradigm that has been verified in computer vision\cite{wangDETR3D3DObject2022, linSparse4DV2Recurrent2023, jiangFar3DExpandingHorizon2023}. 

Additionally, the 'detection by tracking' strategy\cite{panMovingObjectDetection2023} is also noteworthy. By leveraging Doppler information, a key feature distinguishing mmWave radar from LiDAR, dynamic information like scene flow can be estimated first. Temporal information can then be used for feature denoising and enhancement, followed by object detection head. We believe this paradigm has the potential to improve multi-object detection and tracking accuracy in complex dynamic autonomous driving scenarios.

\subsection{Specialized Information Utilizing}
Compared with LiDAR, the 4D mmWave radar is characterized by its specialized information, such as pre-CFAR data and Doppler information. Their comprehensive utilization holds great importance in the competition between 4D mmWave radar and LiDAR technologies.

\subsubsection{Pre-CFAR Data}
Regarding the distinctive data formats throughout the 4D mmWave radar signal processing workflow before CFAR, such as raw ADC data, RD maps, and 4D tensors, their utilization for perception and SLAM tasks represents an interesting yet largely unexplored area of research. The development of learning-based models that can effectively leverage the information contained within these data formats, while maintaining satisfactory real-time performance, could potentially emerge as a focal point in future research endeavors.

\subsubsection{Doppler Information}
The velocity measurement ability based on Doppler effect makes the 4D mmWave radar a unique sensor. Separating point clouds from static and dynamic empowers applications such as object detection, semantic segmentation, localization, and so on. Existing research has already made certain explorations into the utilization of Doppler information, but there are still great potential of it. For example, Doppler-based multiple object tracking may achieve higher accuracy than traditional methods. Besides, Doppler information is also an indispensable feature in network designing. Related architectures such as Doppler-based attention can promote better feature extraction of the scene.


%


\subsection{Dataset Enriching}
As with all other data-driven research domains, datasets pertaining to 4D mmWave radars play a significant role in facilitating related studies. Though as Section \ref{Sec:Datasets} illustrates, there are already several datasets dedicated in 4D mmWave radar, the scale of each dataset and the standardization of 4D mmWave radars are still two main concerns about dataset.

\subsubsection{Scale Expansion}
The largest 4D mmWave radar dataset contains only 152K labeled frames \cite{madaniRadatronAccurateDetection2022}. As a comparison, there are 1.4M and 7M frames of image in nuscenes\cite{caesarNuScenesMultimodalDataset2020} and ONCE \cite{maoOneMillionScenes2021} dataset, respectively. To ensure the generalizability of 4D mmWave radar algorithms, dataset scale expansion is non-negligible.

\subsubsection{Standardization}
The 4D mmWave radar is industry just emerging recently, with many manufacturers being newly established startups. This brings about the problem of the type standardization of 4D mmWave radars. Existing datasets contains varying types of 4D mmWave radars with different parameters such as angular resolution and the largest detection range, which hinders the cross testing in different datasets. Therefore, the type standardization of 4D mmWave radars is also of great necessity.

\subsection{Tasks Exploring}


Although the integration of 4D mmWave radar into autonomous driving systems has shown promising advancements, we still find certain critical applications have not yet been extensively explored.

\subsubsection{Scene Reconstruction and Generation}


Scene reconstruction and generation are pivotal in synthesizing realistic scenarios from actual vehicular data, allowing for object manipulation within these scenarios to generate new data sets. These techniques are instrumental in producing a substantial amount of realistic data, accelerating testing cycles, and greatly alleviate the long-tail data challenges faced by autonomous driving. In recent years, numerous algorithms for scene reconstruction and generation tasks based on Neural Radiance Field(NeRF)\cite{mildenhall2021nerf} or 3D Gaussian Splatting\cite{kerbl20233d} have emerged within the visual domain, but research integrating these methods with 4D mmWave radar remains sparse.  The primary challenges in adapting these visual-based methods to mmWave radar technology include their lack of sensitivity to the electromagnetic wave reflection properties of different materials, resulting in discrepancies between the synthetically rendered point clouds and those captured by actual mmWave radar sensors.

\subsubsection{4D Occupancy Prediction}


Occupancy Prediction is an emerging task aimed at detecting irregularly-shaped and out-of-vocabulary objects, providing detailed occupancy states and semantic information for each spatial grid. However, existing methods based on vision or LiDAR \cite{tianOcc3DLargeScale3D2023, wangOpenOccupancyLargeScale2023, weiSurroundOccMulticamera3D2023} lack consideration of target velocity, making them difficult to apply to downstream decision-making and planning processes in autonomous driving. The incorporation of the Doppler effect, a feature inherent to mmWave radar, presents a novel opportunity. By harnessing the Doppler dimension, 4D mmWave radar could simultaneously estimate occupancy, semantic attributes, and velocity of objects within spatial grids, offering a comprehensive scene discription that could significantly enhance the performance and reliability of autonomous driving systems.

\section{Conclusion} \label{Sec:Conclu}
This paper offers a comprehensive overview of the role and potential of 4D mmWave radars in autonomous driving. It sequentially delves into the background theory, learning-based data generation methods, application algorithms in perception and SLAM, and related datasets. Furthermore, it casts a forward-looking gaze towards future trends and potential avenues for innovation in this rapidly evolving field. The exploration of 4D mmWave radars within the scope of autonomous driving is an ongoing endeavor. This comprehensive review serves as both a primer for those new to the field and a resource for experienced researchers, offering insights into the current state of the art and highlighting the potential for future developments.


\bibliographystyle{IEEEtran} 

\bibliography{4dRadar}

\begin{IEEEbiography}
[{\includegraphics[width=1in,height=1.25in,clip,keepaspectratio]{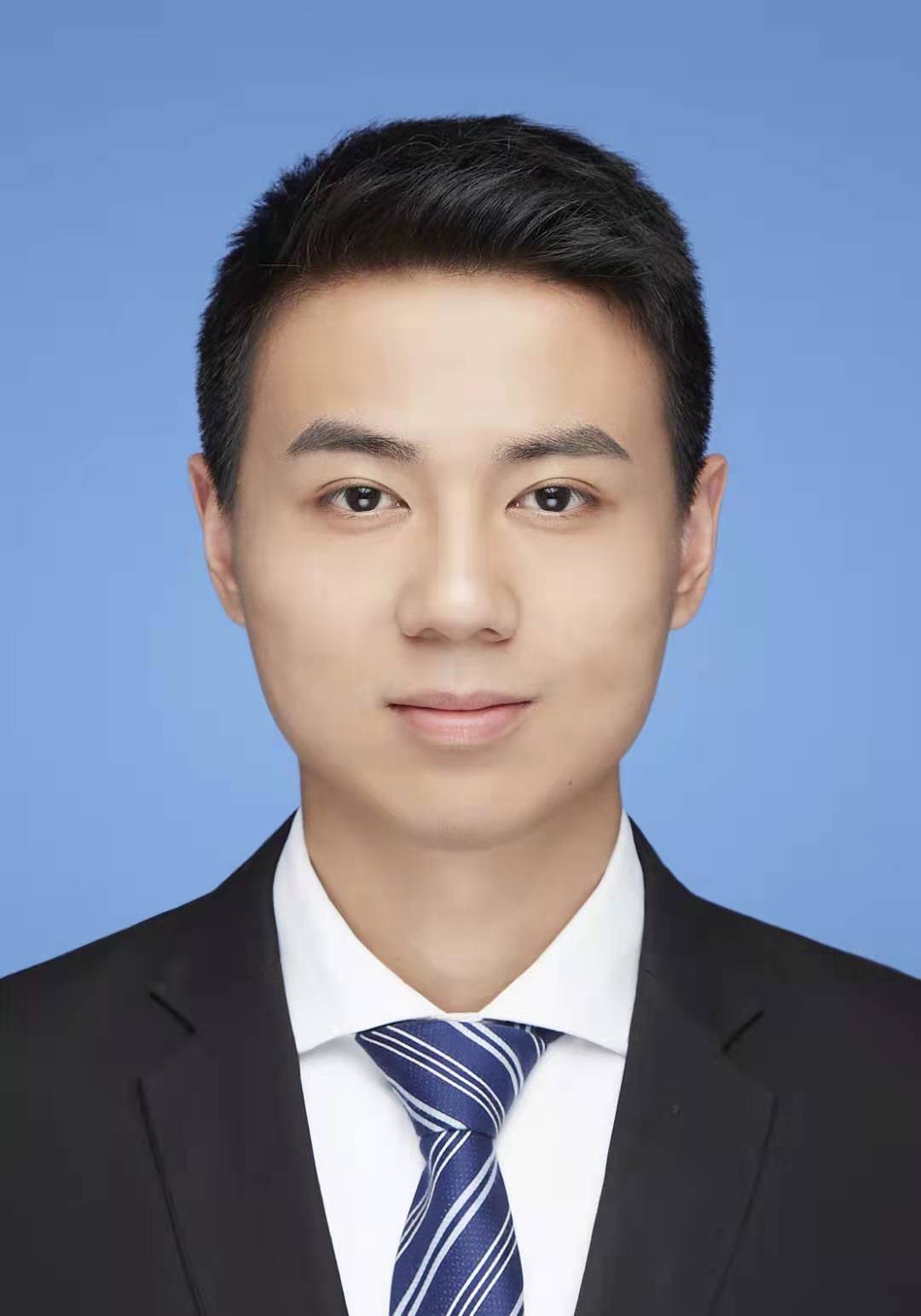}}]
{Zeyu Han} 
received the bachelor’s degree in automotive engineering from School of Vehicle and Mobility, Tsinghua University, Beijing, China, in 2021. He is currently pursuing the Ph.D. degree in mechanical engineering with School of Vehicle and Mobility, Tsinghua University, Beijing, China. His research interests includes autonomous driving SLAM and environment understanding by 4D mmWave radars.
\end{IEEEbiography}

\begin{IEEEbiography}
[{\includegraphics[width=1in,height=1.25in,clip,keepaspectratio]{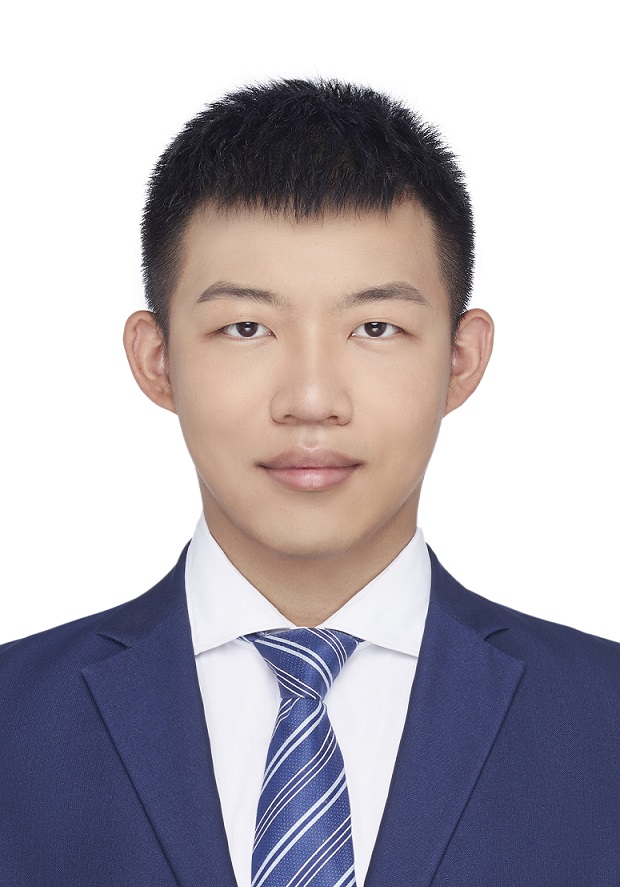}}]
{Jiahao Wang} 
earned the Bachelor's degree in Automotive Engineering from the School of Vehicle and Mobility at Tsinghua University, Beijing, China, in 2022. He is presently pursuing the Ph.D. degree in Mechanical Engineering at the same institution. His current research is centered on scene comprehension and robust multimodal 3D perception for autonomous driving.
\end{IEEEbiography}

\begin{IEEEbiography}
[{\includegraphics[width=1in,height=1.25in,clip,keepaspectratio]{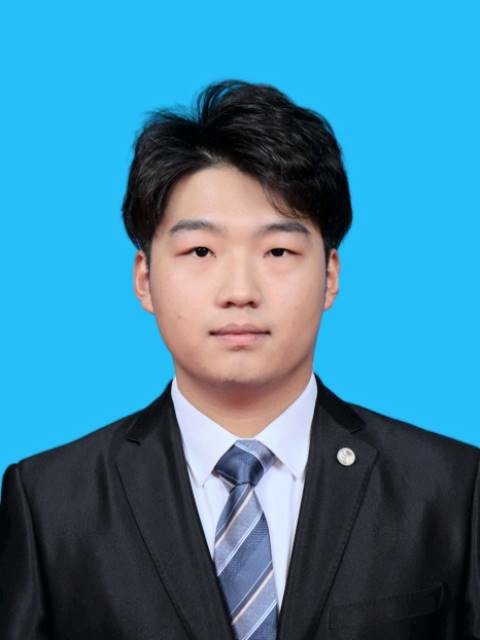}}]
{Zikun Xu} 
earned his bachelor's degree in measurement and control technology and instruments from the School of Mechanical, Electronic and Control Engineering, Beijing Jiaotong University, Beijing, China, in 2023. He is currently pursuing his Ph.D.degree in mechanical engineering with the School of Vehicle and Mobility, Tsinghua University, Beijing 100084, China. His research interests include drivable area detection and scene understanding.
\end{IEEEbiography}

\begin{IEEEbiography}
[{\includegraphics[width=1in,height=1.25in,clip,keepaspectratio]{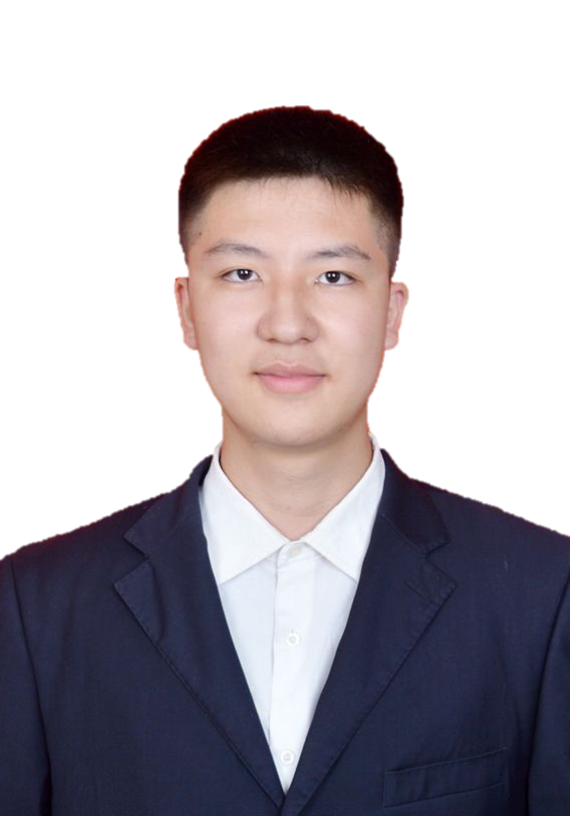}}]
{Shuocheng Yang} 
is currently pursuing a bachelor's degree in Theoretical and Applied Mechanics and Vehicle Engineering at Tsinghua University's Xingjian College. His research focuses on radar-based SLAM.
\end{IEEEbiography}

\begin{IEEEbiography}
[{\includegraphics[width=1in,height=1.25in,clip,keepaspectratio]{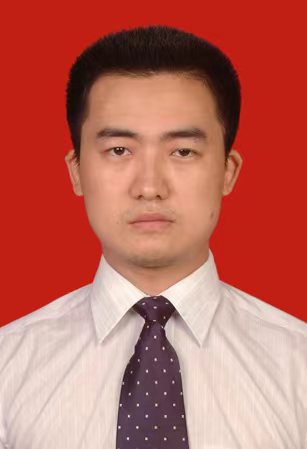}}]
{Zhouwei Kong} 
is currently working at Changan Automobile as a senior engineer, primarily focusing on the mass production and deployment of advanced driver assistance systems (ADAS). He established the decision and control team at Changan and led the development of Changan's first-generation pilot assistance system.
\end{IEEEbiography}

\begin{IEEEbiography}
[{\includegraphics[width=1in,height=1.25in,clip,keepaspectratio]{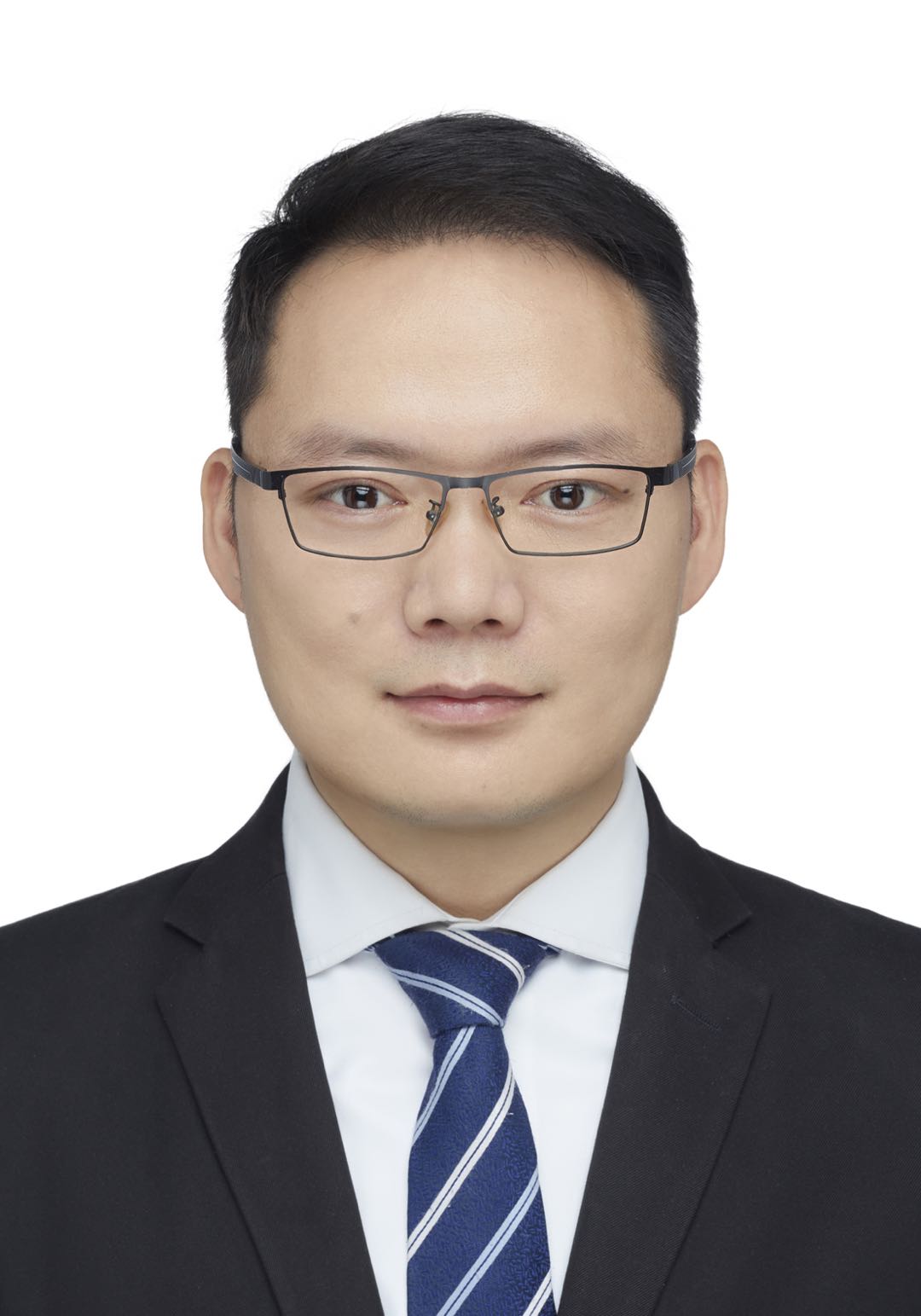}}]
{Lei He} 
received his B.S. in Beijing University of Aeronautics and Astronautics, China, in 2013, and the Ph.D. in the National Laboratory of Pattern Recognition, Chinese Academy of Sciences, in 2018. From then to 2021, Dr. He served as a postdoctoral fellow in the Department of Automation, Tsinghua University, Beijing, China. He worked as the research leader of the Autonomous Driving algorithm at Baidu and NIO from 2018 to 2023. He is a Research Scientist in automotive engineering with Tsinghua University. His research interests include Perception, SLAM, Planning, and Control.
\end{IEEEbiography}

\begin{IEEEbiography}
[{\includegraphics[width=1in,height=1.25in,clip,keepaspectratio]{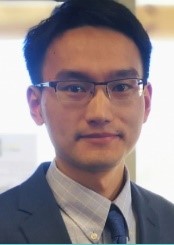}}]
{Shaobing Xu} 
received his Ph.D. degree in Mechanical Engineering from Tsinghua University, Beijing, China, in 2016. He is currently an assistant professor with the School of Vehicle and Mobility at Tsinghua University, Beijing, China. He was an assistant research scientist and postdoctoral researcher with the Department of Mechanical Engineering and Mcity at the University of Michigan, Ann Arbor. His research focuses on vehicle motion control, decision making, and path planning for autonomous vehicles. He was a recipient of the outstanding Ph.D. dissertation award of Tsinghua University and the Best Paper Award of AVEC’2018.
\end{IEEEbiography}

\begin{IEEEbiography}
[{\includegraphics[width=1in,height=1.25in,clip,keepaspectratio]{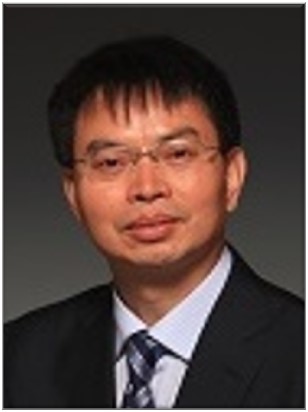}}]
{Jianqiang Wang} 
received the B. Tech. and M.S. degrees from Jilin University of Technology, Changchun, China, in 1994 and 1997, respectively, and Ph.D. degree from Jilin University, Changchun, in 2002. He is currently a Professor of School of Vehicle and Mobility, Tsinghua University, Beijing, China. He has authored over 150 papers and is a co-inventor of over 140 patent applications. He was involved in over 10 sponsored projects. His active research interests include intelligent vehicles, driving assistance systems, and driver behavior. He was a recipient of the Best Paper Award in the 2014 IEEE Intelligent Vehicle Symposium, the Best Paper Award in the 14th ITS Asia Pacific Forum, the Best Paper Award in 2017 IEEE Intelligent Vehicle Symposium, the Changjiang Scholar Program Professor in 2017, Distinguished Young Scientists of NSF China in 2016, and New Century Excellent Talents in 2008.
\end{IEEEbiography}

\begin{IEEEbiography}
[{\includegraphics[width=1in,height=1.25in,clip,keepaspectratio]{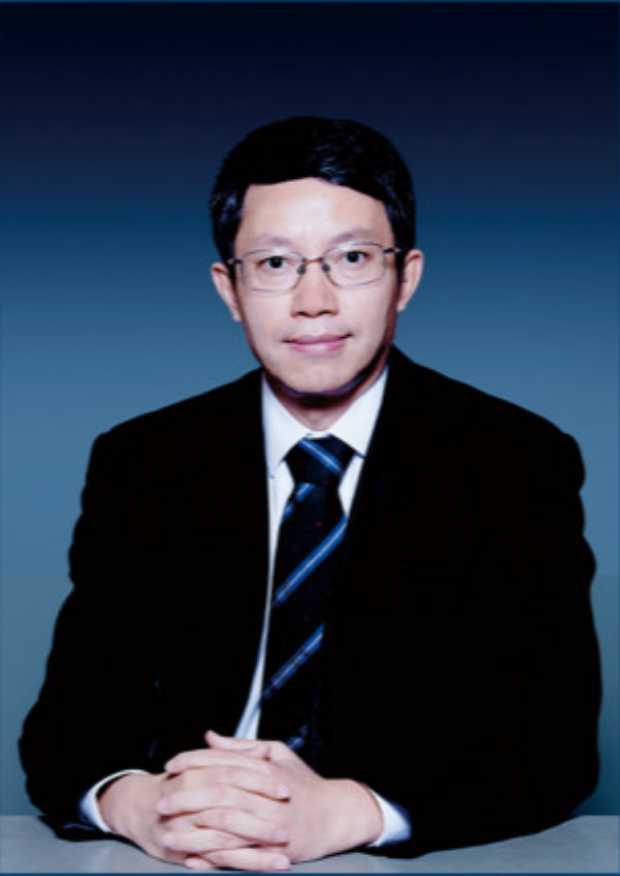}}]
{Keqiang Li} 
received the B.Tech. degree from Tsinghua University of China, Beijing, China, in 1985, and the M.S. and Ph.D. degrees in mechanical engineering from the Chongqing University of China, Chongqing, China, in 1988 and 1995, respectively. He is currently a Professor with the School of Vehicle and Mobility, Tsinghua University. His main research areas include automotive control system, driver assistance system, and networked dynamics and control, and is leading the national key project on CAVs (Connected and Automated Vehicles) in China. Dr. Li has authored more than 200 papers and is co-inventors of over 80 patents in China and Japan. Dr. Li has served as a Fellow Member of Chinese Academy of Engineering, a Fellow Member of Society of Automotive Engineers of China, editorial boards of the International Journal of Vehicle Autonomous Systems, Chairperson of Expert Committee of the China Industrial Technology Innovation Strategic Alliance for CAVs (CACAV), and CTO of China CAV Research Institute Company Ltd. (CCAV). He has been a recipient of Changjiang Scholar Program Professor, National Award for Technological Invention in China, etc.

\end{IEEEbiography}

\addtolength{\textheight}{-12cm}  

\end{document}